\documentstyle[12pt]{article}
\newcommand{\Od}{{\cal O}}
\newcommand{\Ima}{\hbox{Im}}
\newcommand{\Rea}{\hbox{Re}}
\newcommand{\NP}[1]{{\em Nucl.\ Phys.\ }{\bf #1}}

\newcommand{\PL}[1]{{\em Phys.\ Lett.\ }{\bf #1}}

\newcommand{\PR}[1]{{\em Phys.\ Rev.\ }{\bf #1}}
\newcommand{\PRL}[1]{{\em Phys.\ Rev.\ Lett.\ }{\bf #1}}

\begin{document}
\input epsf
\thispagestyle{empty}
\hfill    LBL-38645 

\hfill    UCM-FT 3/96 

\hfill    April 1996

{\footnotesize 

\hfill    Revised Nov96 

\hfill 	  To appear in:

\hfill    PRD 1-Sept-97}

\begin{center}
{\LARGE \bf  The inverse amplitude method in Chiral \\  Perturbation Theory}
\footnote{This work was supported by the Director, Office of Energy
Research, Office of High Energy and Nuclear Physics, Division of High
Energy Physics of the U.S. Department of Energy under Contract
DE-AC03-76SF00098.}

\vskip .5cm

{\Large A. Dobado\normalsize$^2$\\
Departamento de F\'{\i}sica Te\'orica.\\
Universidad Complutense.    
28040 Madrid, SPAIN\\
and

\vskip .3cm

\Large J.R. Pel\'aez\normalsize$^{3}$ \\
Theoretical Physics Group. Lawrence Berkeley Laboratory\\
University of California. 
Berkeley, California 94720. USA.}

\end{center}

\vskip .5cm

\begin{abstract}
Based on a dispersive approach, we apply the inverse amplitude method to
unitarize one-loop $SU(2)$ and $SU(3)$ Chiral Perturbation Theory.
Numerically, we find that this unitarization technique
yields the correct complex 
analytic structure in terms of cuts and poles.
Indeed, using the chiral parameter
estimates obtained from low energy experiments
 we obtain the  poles associated with
the $\rho(770)$ and $K^*(982)$ resonances.
Just by fixing their actual masses
 we obtain a parametrization of the $\pi\pi$ and $\pi
K$ phase shifts in eight different channels.
 With this fit we have then calculated several low-energy 
phenomenological parameters estimating their errors.
Among others, we have obtained the chiral parameters
and scattering lengths, which 
can be relevant for future experiments.
\end{abstract}

\footnotetext[2]{E-mail:dobado@eucmax.sim.ucm.es}
\footnotetext[3]{
Complutense del Amo fellow. On leave of absence from:
Departamento de F\'{\i}sica Te\'orica. Universidad Complutense.
28040 Madrid, Spain. E-mail:pelaez@theorm.lbl.gov, pelaez@vxcern.cern.ch}
\newpage

\section{Introduction}

Even though QCD yields
a remarkably good description of the strong interaction, the low energy
hadron physics has to be modelled phenomenologically. This is due to the fact
that the usual perturbative approach in the coupling constant cannot be applied
to QCD below energies of the order of 1 GeV. Most of the phenomenological
results  were based on PCAC and Current Algebra. However, in 
1979, Weinberg \cite{Weinberg} showed how to reobtain many of these 
predictions by means of an effective lagrangian.

    The fields in that lagrangian are the light mesons, (pions, kaons and
etas) which are understood
as the Goldstone Bosons (GB) arising from the spontaneous breaking of chiral
symmetry. The lagrangian is built as an
expansion in derivatives, that respects the
symmetry breaking pattern of QCD.
 Indeed, the first term in
the expansion is fixed by the symmetry requirements and accounts for the 
Current Algebra results. The next terms in the expansion 
produce further corrections, which depend on several phenomenological 
parameters but are always consistent with the 
QCD symmetry constraints. 
These techniques were later developed to one loop  
in a set of papers by Gasser and Leutwyler \cite{GL1,GL2}. They showed
how to obtain amplitudes involving light mesons,
 as functions of their momenta, their masses and those few
phenomenological parameters.

By fitting these parameters from a few 
low energy experiments it is then
possible to obtain successful predictions for other processes. The
whole approach is 
known as Chiral Perturbation Theory (ChPT). 

Very recently
some partial higher order calculations \cite{Knecht} have appeared 
in the literature
as well as a complete two loop calculation of $\pi\pi$ scattering
\cite{pi2}, which will be needed in order to analyze new data
to come from DA$\Phi$NE and Brookhaven. For a general review of the available
experimental data on pion physics and future prospects, we refer the
reader to \cite{Proceedings}.

    Nevertheless, there are some intrinsic limitations when applying ChPT,
namely, the fact that the amplitudes calculated within the chiral approach
are only unitary in the perturbative sense, that is, up to the next order in the
external momenta. Such a breakdown of unitarity is most severe at high energies,
where the external momenta is no longer a good expansion parameter, 
although it can also occur at 
moderate energies \cite{Truong}. As a result, it is not possible to reproduce
resonant states, which are one of the most characteristic
features of the strongly interacting regime.
Many different methods have been proposed in 
order to improve this behavior and thus to extend the applicability of ChPT to
higher energies; among them: The use of Pad\'e approximants \cite{DoHe}, 
the explicit introduction of resonances \cite{res,nuevo}, 
the K-matrix \cite{KMa}, the large $N$ limit
\cite{largen} (N being the number of GB)
or the inverse amplitude method (IAM) \cite{Truong,DoHe,UpiK,Hannah}. 

This work is devoted precisely to the last method, which can be justified
within a dispersive approach and can easily 
reproduce the two lightest resonances: the $\rho(770)$ in 
$\pi\pi$ scattering \cite{DoHe} 
and the $K^*(892)$ in $\pi K$ scattering \cite{UpiK}. But not only
that, the IAM
also improves considerably the fit to data even in non-resonant
channels, almost up to the first
two particle inelastic threshold (The many
particle inelastic thresholds can be neglected since they are
 suppressed by phase space factors). This fit provides a remarkably good
parametrization that can be used for other processes.
Indeed, in a previous work \cite{photon}, 
the authors showed how it can be used together with
a simple unitarization prescription to obtain successful results on
 $\gamma \gamma \rightarrow \pi^0 \pi^0$ up to 700 MeV.

Of course, it is also possible to obtain very good parametrizations 
\cite{res,nuevo} of $\pi\pi$ or $\pi K$ elastic scattering 
 by including all resonant states
explicitly.
However, our aim choosing
the IAM is to reproduce these phenomena just with
the few phenomenological parameters present in the ChPT lagrangian. In this way,
even though their masses and widths
will not be obtained with great accuracy,
resonances can be regarded as real predictions.
That is one of the relevant features of the IAM since other very popular
unitarization methods are not able to reproduce resonances 
unless they are explicitly introduced in the calculation. 
That is, for instance, the case with the K-matrix.

The purpose of this work is, first, to study how high in energies
 the IAM yields good results and what are its
 limitations. We would also like to
know whether it is possible to reproduce further resonance states. It is clear
that the best candidates are the lightest resonances whose  
dominant decay modes
are $\pi\pi$ or $\pi K$. We have listed them in Table 1. In case these
resonances were not accommodated after our unitarization, it would be interesting
to understand why. Second, once we have a good fit to these resonances, we want
to make a complete numerical analysis of  several low-energy quantities of interest,
like the chiral parameters or the scattering lengths,
including estimations for their errors. 
As we will see below, we expect that the IAM somehow will include effects that cannot
be obtained from the pure $p^2$ expansion.

\begin{table}[h]
\begin{center}
\begin{tabular}{|c|c|c|c|c|} \hline
Name & I,J & Mass & Width & Dominant decays \\ \hline\hline
$\rho(770)$ & 1,1 & 768.8 $\pm$ 1.0 & 150.3 $\pm$ 1.0 & $\pi\pi$, 100\%  \\
\hline
 & & & &$\pi\pi$, (78.1 $\pm$2.4)\%  \\
\raisebox{1.5ex}[1ex]{$f_0(980)$ } &
\raisebox{1.5ex}[1ex]{ 0,0 } &
\raisebox{1.5ex}[1ex]{ 980 $\pm$ 10} &
\raisebox{1.5ex}[1ex]{ 40 to 400} & 
$K \bar K$, (21.9 $\pm$ 2.4)\% \\ \hline 
$f_2(1270)$& 0,2 & 1275 $\pm$ 5 & 185 $\pm$ 20 & $\pi\pi$, (84.7 $\pm$2.6)\% 
\\ \hline \hline
$K^*(892)^\pm$& 1/2,1 & 891.59 $\pm$ 0.24 & 49.8 $\pm$ 0.8 &\\ 
$K^*(892)^0$& 1/2,1 & 896.10 $\pm$ 0.28 & 50.5 $\pm$ 0.6  &
\raisebox{1.5ex}[1ex]{$\pi K$, $\simeq$ 100\%}\\ \hline
\end{tabular}
\end{center}

\leftskip 1cm
\rightskip 1cm

{\footnotesize {\bf Table 1:} 
Lightest resonances with $\pi\pi$ or $\pi K$ dominant decay modes. Data taken
from \cite{PDG}.}

\leftskip 0.cm
\rightskip 0.cm

\end{table}

    Finally, we would like to comment on another motivation of the present work,
which at first may not seem very related to the main topic. 
The philosophy of the chiral approach has also reached
the description of
the strongly interacting symmetry breaking sector (SISBS) of the Standard Model
\cite{SMSBS}.
The scalar sector of such a model displays the same symmetry breaking pattern as
two flavor massless QCD. Hence it is possible to build an effective lagrangian,
much 
as it is done for ChPT \cite{Appel}. Although the electroweak GB are not
physical, using this lagrangian it is possible to obtain predictions for the
scattering of longitudinal gauge bosons \cite{sbseff} at future colliders, like
the LHC. Indeed, there are already experimental proposals to measure the 
electroweak chiral
 parameters at CMS \cite{nosotros}.
Most of the works on the SISBS make use of the
Equivalence Theorem \cite{SMSBS}, which allows us to read the observable 
amplitudes,
 in terms of longitudinal gauge bosons, directly from those with GB.
This theorem has been recently proved in the chiral lagrangian formalism
\cite{ET} and seems to be severely constrained by the lack of unitarity.
At this point is when the unitarization procedures come into play
and it is
crucial to know whether they are
reliable, since what we are now looking for are real predictions and not
elaborated fits to still unavailable data.

In Section 2 we review some basic aspects of exact and perturbative 
unitarity and we define the partial waves in elastic scattering. Section 3 
introduces the IAM, first with a derivation from 
Dispersion Theory and then by studying the constraints to its
 applicability. Section 4 and Section 5 are organized in the 
same way, although they refer to $SU(2)$ and $SU(3)$ ChPT, respectively:
First we apply
 the IAM to ChPT with the chiral parameters obtained from low energy
experiments in order to study the IAM predictive power. Next, we
present an IAM fit to the data. For the best $SU(3)$ fit we present 
 the unitarized results for the 
scattering lengths and some other phenomenological parameters.
 Then, in Section 6, we
study the analytic structure on the complex plane of the IAM amplitudes. In Section
7 we present the conclusions. There is also an Appendix
where we give the elastic scattering formulae used in this work,
as well as a discussion on perturbative unitarity. 

\section{Partial waves, phase shifts and unitarity.}

When dealing with strong interactions, it is usual to project the amplitudes
in partial waves with definite angular momentum $J$ and isospin $I$,
as follows
\begin{equation}
t_{IJ}(s)=\frac{1}{32 K \pi}\int_{-1}^{1}
\mbox{d}(\cos \theta) P_J(\cos
\theta)T_I(s,t)
\end{equation}
where $K=2$ or $1$ depending on whether the particles in the process are
identical or not. The acceptable isospin values also depend on the process,
namely $I=0,1,2$ for $\pi\pi$ elastic scattering and $I=1/2,3/2$ for $\pi K$.
For both reactions the definite isospin amplitudes $T_I$ are obtained 
from a single
function. In the first case
\begin{eqnarray}
T_0(s,t,u)&=& 3 A(s,t,u) + A(t,s,u) + A(u,t,s) \nonumber \\
 T_1(s,t,u)&=& A(t,s,u) - A(u,t,s) \nonumber \\
T_2(s,t,u)&=&A(t,s,u) + A(u,t,s)
\end{eqnarray}
whereas for $\pi K$ scattering we can write
\begin{equation}
T_{1/2}(s,t,u)=\frac{3}{2}T_{3/2}(u,t,s)-\frac{1}{2}T_{3/2}(s,t,u)
\end{equation}

In order to deal with both processes on the same footing, we will label the 
particles in the reaction as $\alpha$ and $\beta$. Thus
the Mandelstam variables will satisfy: $s+t+u=2(M_{\alpha}^2+M_{\beta}^2)$
and the threshold will
be at $s_{th}=(M_\alpha + M_\beta)^2$. As it is well
known, whenever $s>s_{th}$, and below inelastic thresholds, 
the unitarity of the S-matrix implies 
\begin{equation}                            
\Ima t_{IJ}=\sigma_{\alpha \beta} \mid t_{IJ} \mid ^2
\label{uni}
\end{equation}
where $\sigma_{\alpha \beta}$ is 
the two particle phase space. Explicitly:
\begin{equation}
 \sigma_{\alpha\beta}(s) = \sqrt{  \left(1-\frac{(M_{\alpha}+M_{\beta})
 ^2}{s}\right)
\left(1-\frac{(M_{\alpha}-M_{\beta})^2}{s}\right)}
\label{sigma}
\end{equation}
As a consequence of Eq.\ref{uni}, the partial wave can be parametrized as follows:
\begin{equation}
t_{IJ}(s)= \frac{1}{\sigma_{\alpha \beta}(s)}
e^{i\delta_{IJ}(s) }\sin \delta_{IJ}(s)
\end{equation}
and $\delta_{IJ}(s)$ is called the $IJ$ phase shift.

We have already mentioned that the ChPT amplitudes are obtained
as an expansion in external
momenta and masses. That is
\begin{equation}
t_{IJ}\simeq t_{IJ}^{(0)}+t_{IJ}^{(1)}+t_{IJ}^{(2)}+...
\end{equation}
where, for the cases we are interested in, $t_{IJ}^{(0)}$ is $\Od (p^2)$,
$t_{IJ}^{(1)}$ is $\Od (p^4)$, etc... In practice, we can only obtain
the few first terms of the series above and therefore the amplitude
only satisfies the unitarity condition perturbatively
\begin{eqnarray}
\Ima t_{IJ}^{(0)}&=& 0 \nonumber \\
\Ima t_{IJ}^{(1)}&=&\sigma_{\alpha \beta} t_{IJ}^{(0)2} \nonumber \\
\Ima (t_{IJ}^{(2)}+t_{IJ}^{(1)})&=&\sigma_{\alpha \beta} \left( 
t_{IJ}^{(0)2} + 2 t_{IJ}^{(0)}
\Rea t_{IJ}^{(1)}\right) 
\simeq \sigma_{\alpha \beta} \mid t_{IJ}^{(0)} + t_{IJ}^{(1)}\mid^2
\label{punit}
\end{eqnarray}
                   
The $\Od (p^2)$ terms were given by Weinberg 
\cite{Weinberg} and they
are called the low energy theorems. 
The next order contributions to $\pi\pi$ scattering were given in
\cite{GL1,GL2}. The calculation for $\pi K$ scattering can be found in
\cite{Bernard1,Bernard2}, although we have found that the formulae
in the literature do not satisfy Eq.\ref{punit}. We will comment on that later. 
Very recently it has appeared the complete calculation of the
$\Od(p^6)$ contribution to elastic $\pi\pi$ scattering \cite{pi2}.
Although we will not use it, we 
will compare some of its results with those of our method.

\section{The Inverse Amplitude Method}

\subsection{Derivation from Dispersion Theory}

Let us briefly review the standard derivation \cite{Truong,UpiK} of the inverse 
amplitude method, since we will use it later 
in order to understand 
the applicability of the method.

Any partial wave obtained from a relativistic
Quantum Field Theory should present a characteristic analytic structure
in the complex $s$ plane. Indeed, the reaction
threshold  becomes a cut in the
from $s_{th}$ to $+\infty$.
Due to crossing symmetry, there
should be another left cut along the negative axis. If we now apply 
Cauchy's Theorem to our complex amplitudes we obtain integral equations
known as dispersion relations. For instance, a three times subtracted
dispersion relation is
\begin{equation}
t_{IJ}(s)=C_0+C_1s+C_2s^2+
\frac{s^3}\pi\int_{(M_{\alpha}+M_{\beta})^2}^{\infty}\frac{\Ima
t_{IJ}(s')ds'}{s'^3(s'-s-i\epsilon)} + LC(t_{IJ})
\label{disp}
\end{equation}
Where we have not written explicitly the left cut (LC) contribution.
The number of subtractions needed depends on how the amplitude
behaves at infinity in order to ensure the vanishing of the contributions 
coming from closing the integral contour. We have chosen
three subtractions since we are going to use $\Od (p^4)$ ChPT amplitudes 
which at high $s$ behave as $s^2$. But our arguments remain valid for 
$\Od (p^6)$ amplitudes when using four times subtracted dispersion relations,
etc...

The ChPT partial waves present both cuts
and we can calculate 
both the subtraction constants $C_0, C_1, C_2$ 
and the integrand inside
Eq.\ref{disp} {\em perturbatively}
\begin{eqnarray}
t_{IJ}^{(0)} &=& a_0+a_1s   \nonumber \\
t_{IJ}^{(1)} &=& b_0+b_1s+b_2s^2+     
\frac{s^3}\pi\int_{(M_{\alpha}+M_{\beta})^2}^{\infty}\frac{\Ima
t_{IJ}^{(1)}(s')ds'}{s'^3(s'-s-i\epsilon)}+LC(t^{(1)}_{IJ}) 
\label{disp1}
\end{eqnarray}
Where we have expanded
 the subtraction constants in terms of $M_\alpha^2/F_\beta^2$.

The IAM is based on the fact that the function $1/t_{IJ}$
displays the very same analytic structure of $t_{IJ}$, apart from some
possible pole contributions. For later convenience, we will make use of
$G(s)=t_{IJ}^{(0)2}/t_{IJ}$. Notice that we have
multiplied $1/t$ by a real function without singularities; 
thus we keep the same analytic structure and
we can write a very similar
dispersion relation:
\begin{equation}
G(s)=G_0+G_1s+G_2s^2+     \\   \nonumber
\frac{s^3}\pi\int_{(M_{\alpha}+M_{\beta})^2}^{\infty}
\frac{\Ima G(s')ds'}{s'^3(s'-s-i\epsilon)}+LC(G)+PC
\label{Gdisp}
\end{equation}
where $PC$ stands for possible pole contributions.
The advantage of using $G(s)$ is that, using Eqs.\ref{uni} and \ref{punit},
we can calculate exactly the integral
over the right cut (but not on the left, since those equations 
only hold on the elastic cut), as follows:
\begin{equation}
\Ima G=-t_{IJ}^{(0)2}\frac{\Ima t_{IJ}}{\mid t_{IJ}\mid^2}= 
-t_{IJ}^{(0)2}\sigma = -\Ima t_{IJ}^{(1)}
\label{ImG}
\end{equation}
Note that we are denote by $t_{IJ}$ the exact amplitude, which is
unknown, although we know its analytic properties. In contrast, the expressions
for $t_{IJ}^{(0)}$ and $t_{IJ}^{(1)}$, etc... have been calculated explicitly.

As we did before, we can also expand the $G_i$ subtraction coefficients in
powers of $M_{\alpha}^2/F_{\beta}^2$, and then rewrite the dispersion relation
for $G(s)$, which now reads
\begin{eqnarray}
\frac{t_{IJ}^{(0)2}}{t_{IJ}}&\simeq& a_0+a_1s-b_0-b_1s-b_2s^2 \nonumber  \\   
&-&\frac{s^3}\pi\int_{(M_{\alpha}+M_{\beta})^2}^{\infty}\frac{\Ima
t_{IJ}^{(1)}(s')ds'}{s'^3(s'-s-i\epsilon)}-LC(t^{(1)}_{IJ})+PC
\simeq t_{IJ}^{(0)}-t_{IJ}^{(1)}
\label{preIAM}
\end{eqnarray}
where we have approximated $\Ima G \simeq -\Ima t_{IJ}^{(1)}$ on the left cut
and we have neglected the
pole contribution. In other words,
\begin{equation}
t_{IJ}\simeq
\frac{t_{IJ}^{(0)2}}{
t_{IJ}^{(0)}-t_{IJ}^{(1)} }
\label{IAM}
\end{equation}
This is the IAM result
that we are going to use in the present work.
Incidentally, Eq.\ref{IAM} can be understood as the formal $[1,1]$
Pad\'e approximant of the ChPT amplitude.

It is important to remark that if we expand again Eq.\ref{IAM}
at low energies, we find
\begin{equation}
t_{IJ}\simeq
\frac{t_{IJ}^{(0)2}}{
t_{IJ}^{(0)}-t_{IJ}^{(1)} }\simeq t_{IJ}^{(0)}+t_{IJ}^{(1)}+{\cal O}(p^6)
\label{recover}
\end{equation}
That is, we recover the ChPT expansion. Hence, up to ${\cal O}(p^6)$
our method and ChPT yield the same low energy results if the same chiral
lagrangian parameters are used.

\subsection{The applicability of the Inverse Amplitude Method}

Let us review all the approximations made in the 
previous section, in order to comment how they will constraint the
IAM applicability:

\subsubsection  {The left cut} 

In Eq.\ref{preIAM} we have replaced
the $G(s)$ left cut integral by that of $-t_{IJ}^{(1)}(s)$. As
we have remarked
 in the preceeding discussion, Eqs.\ref{uni} and \ref{punit} are only exact
on the right cut. On the left cut we cannot write the chain of
equalities that lead to Eq.\ref{ImG}.
Nevertheless, if we use the ChPT result as an approximation:
\begin{equation}
\Ima G=-t_{IJ}^{(0)2}\frac{\Ima t_{IJ}}{\mid t_{IJ}\mid^2}\simeq  -\Ima t_{IJ}^{(1)}
+\Od(p^6)
\label{ImGleft}
\end{equation}
we get
\begin{equation}
LC(G)=\int_{-\infty}^{0}\frac{\Ima
G_{IJ}(s')ds'}{s'^3(s'-s-i\epsilon)}\simeq
- \int_{-\infty}^{0}\frac{\Ima
t_{IJ}^{(1)}(s')ds'}{s'^3(s'-s-i\epsilon)}= - LC(t^{(1)}_{IJ})
\end{equation}
Notice that, in order to obtain the $IJ$ phase shifts, we are going to calculate
 $t_{IJ}(s)$ for real $s>4 M_\pi$. That means that the denominator
 $(s'-s-i\epsilon)$ inside the integrals is never going to be very small, 
which somehow will wash out the error on the left cut.
But note also that treating 
differently the right and left cuts violates crossing symmetry.

Indeed, in \cite{GassMess} it has already been pointed out that the Pad\'e
approximants do not reproduce correctly the subleading
logarithms that would appear at
{\em next order} in the chiral expansion (${\cal O}(p^6)$ in this case).
Of course they would be obtained if we applied the IAM to the chiral
amplitudes at ${\cal O}(p^6)$, but still the method would not
yield the correct logarithms at ${\cal O}(p^8)$ and so on.
At high energies chiral logarithms are not so relevant, 
but at low energies
they are a very important feature of 
ChPT and indeed they can give the dominant
contribution in some channels. 

Nevertheless, from
Eq.\ref{recover} we see that at low energies the IAM yields
 the very same ${\cal O}(p^4)$ ChPT expansion, {\em including} the 
{\em dominant} chiral logarithms. The contribution from the left cut
and subleading logarithms is ${\cal O}(p^6)$.
As a consequence, if we try to make a low energy 
fit to the data, the parameters that we would obtain with the IAM 
would not lie very far from those of ChPT, but they will 
not be the same. That is the reason why, in the following sections,
we will denote with a hat the parameters obtained from any
IAM fit.\footnote{ While we were revising this paper it has appeared
a work by M.Boglione and M.R.Pennington 
\cite{BoPe} where they propose 
other schemes with better approximations to the left cut
and also include possible contributions from Adler zeros.}

\subsubsection  { Resonances and the pole contribution} 

In Eq.\ref{preIAM}, we have neglected the contributions coming from zeros in
the amplitude, that will appear as poles of the inverse function.
There is no way to know
{\em a priori} whether or not a partial wave will vanish 
for a given value of $s$, although
it is known that chiral amplitudes have zeros
below threshold, which are known as Adler zeros.
Their position is not known except for the $I=1,J=1$ 
channel, where the pole is located at threshold. In our 
derivation it is 
compensated by the same zero in the $t^{(0)}_{11}$
channel. That is not the case of the $J=0$ amplitudes
and therefore we are neglecting the contribution of their residue.
Consequently, our amplitudes are not valid to
obtain Adler zeros 
and that will affect our results at low energy 
(but no more than ${\cal O}(p^6)$).
That is another reason 
to differentiate the parameters 
obtained from our fit from 
those of the pure chiral expansion\footnotemark[2].

\subsubsection { Multiplying by $t_{IJ}^{(0)}$}

 This is apparently a harmless
assumption in the above reasoning, although it  
dramatically affects the results of the IAM.
In fact, it can happen that $t_{IJ}^{(0)}=0$. 
In the 
$(I,J)=(0,0),(1,1),(2,0)$ channels of $\pi\pi$ scattering or in the
$(3/2,0),(1/2,0),(1/2,1)$ in $\pi K$, this only occurs for isolated values
of $s$, at or below threshold. In particular, that means that the IAM 
amplitudes will
have the same zeros as the lowest order Chiral amplitudes. 
However, every other partial wave vanishes
at $\Od (p^2)$, for any $s$. As a consequence,
the formula in Eq.\ref{IAM} is no longer valid.

Nevertheless, we can
generalize our previous derivation, in order to
include those channels whose leading order is $\Od (p^4)$.
We only have to go through the very same steps,
although now we would write a dispersion relation for $t_{IJ}^{(2)}$.
But let us remember that the main improvement of the approach is that we are
calculating exactly the integral of $\Ima G(s)$ over the right cut.
However, for that purpose we need an imaginary part, and by looking
 at
Eq.\ref{punit} we can see that $t_{IJ}^{(0)}=0$ implies that $\Ima t_{IJ}^{(1)} =
\Ima t_{IJ}^{(2)}= 0$. Therefore,
unless we have a calculation up to $\Od (p^8)$, the corresponding
imaginary part will vanish. Hence when following
 the derivation of the IAM
if $t_{IJ}^{(0)}=0$ the best we can get is plain ChPT again.
At present, only ${\cal O} (p^6)$ calculations are available and
we can only expect to obtain a real improvement with
our approach in the six channels listed above. 
Thus, we will not be able to reproduce 
the $f_2(1200)$ resonance. 

\subsubsection{Elastic unitarity} 

In order to obtain $\Ima G$ on the right cut, Eq.\ref{ImG}, we have 
just made use of the elastic
unitarity condition of Eq.\ref{uni}.
However,
the right cut is composed of many superimposed cuts, each one 
corresponding
to a different inelastic intermediate channel. Actually, Eq.\ref{uni}
is only true below the first inelastic threshold, and the real unitarity
condition reads
\begin{equation}
\Ima t_{\alpha\beta\rightarrow\alpha\beta} = \sum_A \sigma_A
 \mid t_{\alpha\beta\rightarrow A} \mid ^2 \Theta(s-s_A)
\label{inelunit}
\end{equation}
The sum is over all the physically accessible 
intermediate states $A$, whose phase space is 
$\sigma_{\alpha\beta}$.

As far as we are neglecting electromagnetic interactions,
the first inelastic channel in $\pi\pi$ is the four pion intermediate state,
at 550 MeV.
Similarly, for $\pi K$ is $\pi K \pi\pi$, whose threshold is $\simeq$ 910 MeV.
Strictly speaking, only for lower energies the elastic 
approximation is exact. 
Nevertheless, the contribution of these intermediate states is 
strongly suppressed  by
the four particle phase space and we expect the
IAM to provide a good approximation. 

Unfortunately within the range of energies we
are interested in, 
there are intermediate channels
which are not suppressed 
by phase space. Indeed, at approximately $985$ MeV
the inelastic $K\bar K$ threshold opens up.
Its phase space factor is the $\sigma_{\alpha\beta}$
in Eq.\ref{sigma}, with 
$M_\alpha=M_\beta=M_K$. Therefore, above the two 
kaon threshold we have to reconsider the derivation of the IAM.
 Let us illustrate with $\pi\pi$ scattering
 how inelastic effects modify our result.
    
As the starting point, for $s>s_{K\bar K}$, we have a new unitarity relation:
\begin{equation}
\Ima t=\sigma_{\pi\pi} \mid t \mid ^2 + 
\sigma_{K\bar K} \mid t_K \mid ^2
\label{unitKK}
\end{equation}
where we have denoted by $t$ the generic $t_{IJ}$ pion elastic 
scattering amplitude and by $t_K$ the $IJ$ partial wave of
the process $\pi\pi \rightarrow K\bar K$. Thus we now have,
for $s>s_{K \bar K}$, that 
\begin{equation}
\Ima G=-t_{IJ}^{(0)2}\frac{\Ima t_{IJ}}{\mid t_{IJ}\mid^2}= 
-t_{IJ}^{(0)2}\left( \sigma_{\pi\pi} +\sigma_{K\bar K} \frac{\mid t_K\mid^2}
{\mid t \mid ^2} \right) 
\end{equation}
which differs from Eq.\ref{ImG} in the
 term coming from two kaon intermediate production. If we follow
the very same steps of our previous derivation, we arrive at
\begin{equation}
\frac{t^{(0)2}}{t_{IJ}}\simeq t^{(0)}-t^{(1)}-
\frac{s^3}\pi\int_{4M_\pi^2}^{\infty}\frac{\sigma_{K \bar K}}
{s'^3(s'-s-i\epsilon)} 
\underbrace{
\left( t^{(0)2}(s')
\frac{\mid t_K(s')\mid^2}{\mid t(s')\mid^2}- t_K^{(0)2}(s')
\right) 
}_{\Delta(s')}ds'
\end{equation}

Notice that, using ChPT, $\Delta(s')\simeq 0 + \Od(p^6)$.
But at these high energies that is not negligible. Besides,
we are interested in the above integral for physical values
of $s$ and therefore the denominator will be almost divergent
for some $s'$. For these reasons we cannot neglect this integral
and then we should not trust the IAM since it
could miss some relevant physical features.

That is indeed the case in pion scattering
since, as it can be seen in Table 1, there is one
resonance, the $f_0(980)$, whose nature is closely related to the $K\bar K$
threshold. Nowadays, the interpretation of that
resonance is still controversial:
different authors propose different poles
(not always just one) in the vicinity of the $K\bar K$ inelastic cut
\cite{f0,Zou}.
As we will see later, our approach is not able to reproduce any
of these poles, consistent with the fact that the IAM
makes use just of elastic unitarity.

At this point we want to remark the importance of understanding 
why and when the method does no longer yield the right results.
Let us remember that we are also thinking in possible applications of
this unitarization procedures to the electroweak chiral effective lagrangian,
whose reference model is the Standard Model with a heavy 
Higgs. In such case, one would expect to see a broad resonance in the 
scalar channel and we want to have a unitarization procedure
whose predictions we can trust.

\subsubsection {\bf $\Od (p^4)$ approximation}

 Throughout the derivation of the IAM
 we have been using the chiral amplitudes up to $\Od (p^4)$.
Nevertheless, it is possible to extend the argument to include
higher order terms, as for instance the $\Od (p^6)$ contributions. 
In that case we would have started from a four times subtracted
dispersion relation for the two-loop calculation. Once more, the integral 
over the right cut would be related to the one for $G(s)=t_{IJ}^{(0)2}/t_{IJ}$.
Working out the expansion of the subtraction constants, we would then arrive to
\begin{equation}
 t_{IJ}\simeq \frac{t_{IJ}^{(0)2} } {
t_{IJ}^{(0)}-t_{IJ}^{(1)}+t_{IJ}^{(1)2}/t_{IJ}^{(0)}-t_{IJ}^{(2)}  }
\end{equation}
Again that is the formal [1,2] Pad\'e approximant, and it
satisfies the elastic unitarity condition. 

    As we have already mentioned
two recent papers have appeared with $\Od (p^6)$ calculations of $\pi\pi$ 
scattering within $SU(2)$ ChPT \cite{Knecht,pi2}. 
We have not used these results, since, as we have just seen, they
will not help us to overcome any of the preceeding objections to the IAM.
 However it is quite likely that, have we used them, the 
parameters of the fits that we will present in the next sections would had
been slightly modified.

\section{$\pi\pi$ scattering in $SU(2)$ ChPT}

The inverse amplitude method was first applied \cite{DoHe,photon}
 to $\pi\pi$ scattering without the strange quark.  
In that case, the massless limit displays 
spontaneous symmetry breaking from $SU(2)_L\times SU(2)_R$ to $SU(2)_{L+R}$,
which is nothing but the usual isospin. The $\Od (p^4)$ expression for $\pi\pi$
scattering was obtained in \cite{GL1,Lehman}, and it is written in terms of
four phenomenological parameters $\bar l_1,\bar l_2,\bar l_3, \bar l_4$ 
as well as the
mass and pion decay constants, $M_\pi$ and $F_\pi$.
In this section we will
review how the method is able to reproduce the $\rho$ resonance.
We will show some results for 
recently proposed new parameters in order to test the IAM predictive
power, but we will also present a unitarized fit to the data.
As a novelty we will use not only the
$J=0$ phase
shifts, but also those with $J=2$, in order 
to obtain the best fit with the IAM.
In this new calculation, we have also estimated the error bars of the 
unitarized parameters.

\subsection{Results using low-energy parameters}

Let us now illustrate what happens if we apply the IAM
on the ChPT amplitudes using 
the chiral parameters obtained from low energy
 experiments. We want
to see quantitatively to what extent the main 
physical features are reproduced.

In order to simplify the comparison with previous
works, we have chosen $M_\pi=139.57$ MeV and $F_\pi=93.1$ MeV.
The values of the chiral parameters are not so clear, since they
have considerable error bars. In Table 2 we have listed the different
sets of parameters that we have 
taken from the literature to obtain Fig.1. 

\begin{table}[h]
\begin{center}
\begin{tabular}{|c|c|c|c|} \hline
\rule[-3mm]{0mm}{8mm}Method & $\bar l_1$ & $\bar l_2$ & $M_\rho$\\ \hline\hline
ChPT & -0.62$\pm$0.94 & 6.28$\pm$0.48 & No resonances \\ \hline \hline
Inverse & -0.62$\pm$0.94 & 6.28$\pm$0.48 & 715 MeV \\ \cline{2-4}
Amplitude & -1.7$\pm$1.0 & 6.1$\pm$0.5 & 675 MeV\\ \hline
\end{tabular} 
\end{center}

\leftskip 1cm
\rightskip 1cm
\vskip .2cm
{\footnotesize {\bf Table 2:}
Sets of parameters and methods used in the text.
Those in the first two lines come from $K_{l4}$ decays \cite{Rigg}.
Those in the third, from data on $K_{l4}$ and $\pi\pi$ together with 
some 
unitarization procedure ref.\cite{BiGa}. 
$M_\rho$ is calculated with the central values.}

\leftskip 0.cm
\rightskip 0.cm
\end{table}

Let us remark at this point that  for the ChPT
phase shifts we are using
 the  definition $\delta \simeq \sigma(t^{(0)}+\mbox{Re}t^{(1)})$
suggested  in \cite{GassMess}. Of course, ChPT is just a low energy 
approach, but incidentally, these phase shifts
coincide with those obtained from the K-matrix unitarization
defined as 
\begin{equation}
t^K=\frac{t^{(0)}+\mbox{Re}t^{(1)}}{1-i\sigma(t^{(0)}+\mbox{Re}t^{(1)})}
\label{Kmatrix}
\end{equation}
It can be easily verified that $t^K$ satisfies elastic unitarity,
Eq.\ref{uni}, exactly.
Consequently, the dotted lines in Fig.1
not only give the ChPT predictions,
but also the results of the K-matrix
unitarization. We will thus confirm that such a method is not able 
to reproduce resonances by itself. They have to be added by hand.

In Fig.1 it can be clearly seen, in the $I=1,J=1$ channel,
that the IAM yields a $\rho$-like resonance.
The value of its
mass is obtained from the point where  
$\delta=90^0$ and it lies 10\% to 15\% away
from its real value. In this way, the {\em existence} of the
$\rho$ resonance can be 
regarded as a prediction of the IAM
with ChPT and the parameters obtained from some low energy data.

 It is also evident that the fit of the $I=2,J=0$
channel is correct up to much higher energies.
In Table 2 we have also included the 
values of $M_\rho$ corresponding to each choice of parameters.
For all the cases we have set $\bar l_3=2.9$ and $\bar l_4=4.3$ following reference 
\cite{GL1}.

The only feature of $\pi\pi$ scattering that is evidently missing from
the unitarized results is the $f_0(980)$ resonance in the $I=0,J=0$
channel. In the previous section we saw that 
 this fact is connected with the failure of the
whole approach to reproduce the kaon inelastic cut. But let us
first obtain a better fit to the data. 

\subsection{Unitarized fit}

Now that we have an amplitude that describes the 
right cut, while keeping at the same time the correct polynomial form
from ChPT, it seems natural to use 
$M_\rho$ \cite{DoHe,photon} to fit the data.
 Note that fixing the correct mass does not
imply a good fit. 
\begin{figure}

\vspace{-0.8cm}

\leftskip -2cm
\begin{center}
\mbox{\epsfysize=8.7cm\epsffile{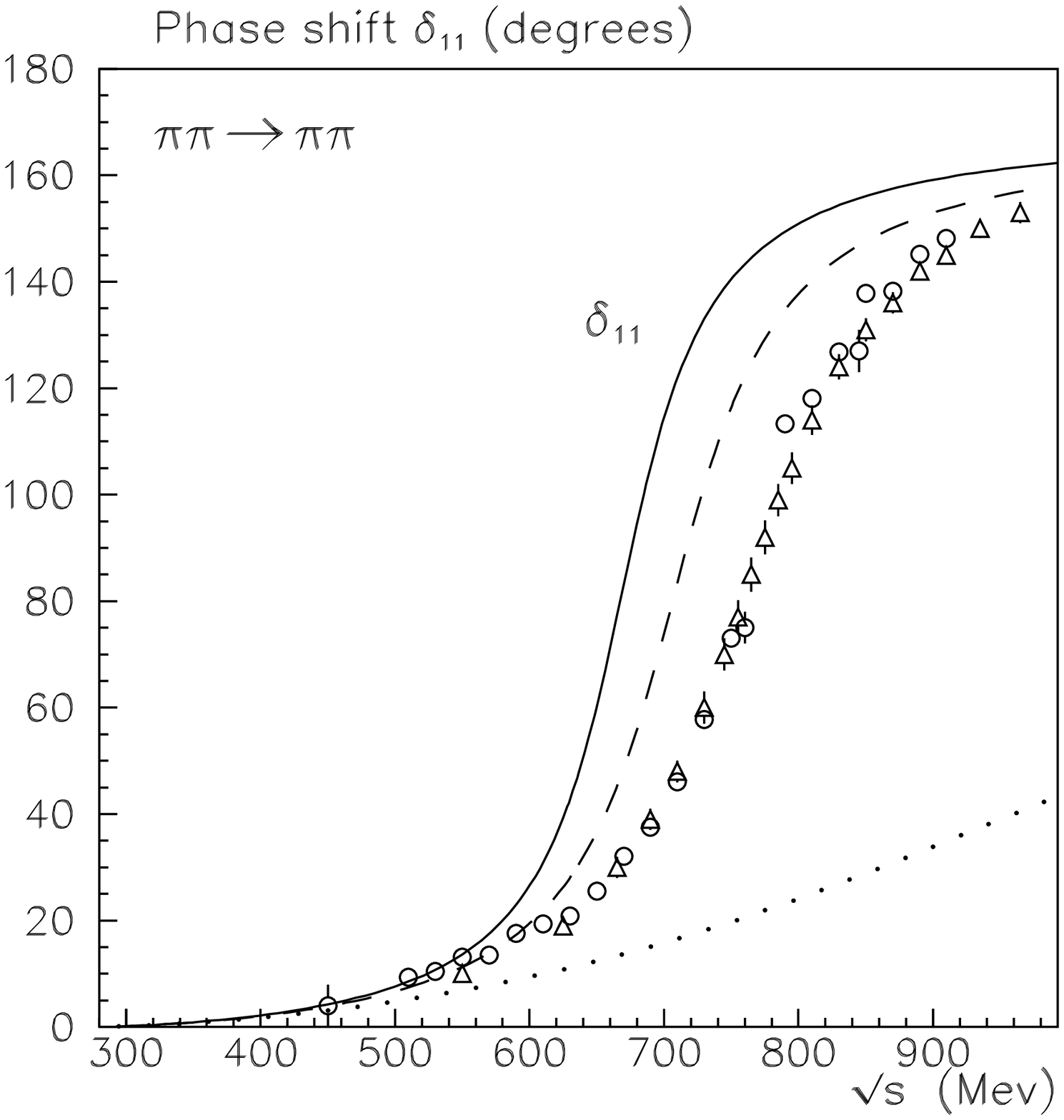}
\epsfysize=8.7cm\epsffile{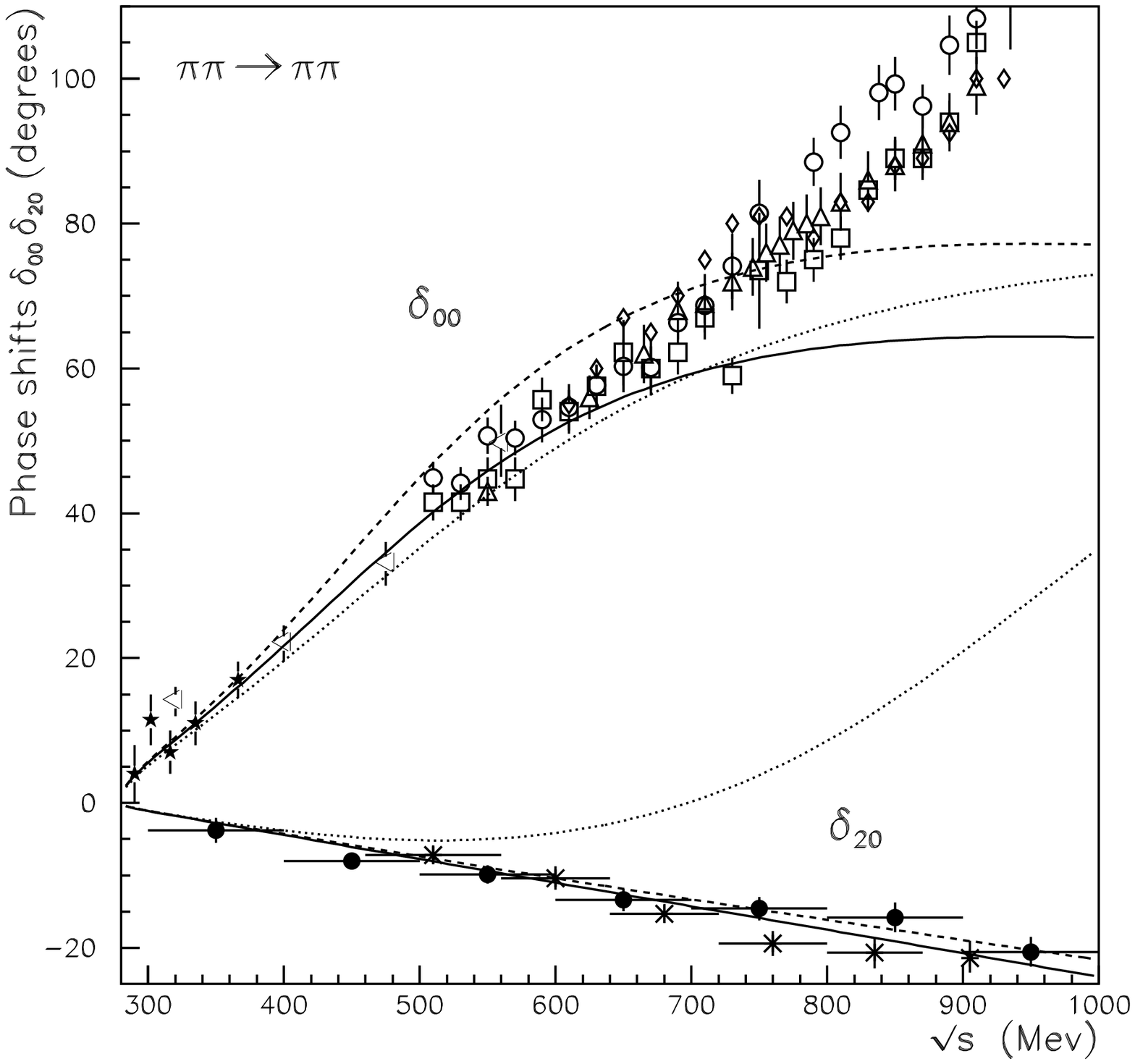}}
\end{center}

\vspace{-0.8cm}

\leftskip 1cm
\rightskip 1cm
{\footnotesize {\bf Figure 1.-} Phase shifts for $\pi\pi\rightarrow\pi\pi$.
The dotted curve is plain ChPT with the $\bar l_i$ in the first
line of Table 2. The other two curves are both the result of the IAM:
the dashed one has been calculated again with the same parameters
whereas the continuous one corresponds to the $\bar l_i$ in
the third line of Table 2. The data come from:
\cite{Proto} ($\triangle$), \cite{Grayer}
($\diamondsuit, \Box$), \cite{Losty}
($\times$),\cite{Esta} ($\circ$), \cite{Srini} ($\lhd$), \cite{Rosselet}
($\star$) and \cite{Hoogland} ($\bullet$).
The results with $SU(3)$ ChPT would have been exactly
superimposed on these curves.
The straight line stands at $\delta=90^0$.}

\leftskip 0.cm
\rightskip 0.cm
\end{figure}
For instance, we could get a wrong width. 
In order to differentiate the parameters thus obtained from
those coming from plain ChPT we will call them $\hat l_1,\hat l_2$.

The (1,1) channel is almost only sensible to
$\bar l_1 - \bar l_2$. With $\hat l_1- \hat l_2=-5.95\pm 0.02$
we get the $M_\rho$ listed in Table 1 and in
Fig. 2 it can be seen that the
results are remarkably successful.
Later we will show that we also get the right width.

Once that difference is fixed, we just have to determine
 one parameter, say $\hat l_2$.
In previous studies \cite{DoHe,photon} the unitarized fit to the 
other phase shifts 
was used in order to estimate the values of $\hat l_1$ and $\hat l_2$. 
But, as we commented above, the data in the (0,0) channel 
is not as good as
that of (1,1). The same happens for the (2,0) channel, where the curves are not
very sensible to small variations in the $\hat l_i$ parameters. Therefore,
in the present work,
we have also used the $J=2$ channels (mainly that with $I=0$)
to further constrain the parameter range.
Let us remember now that in these channels
$t_{I2}^{(0)}=0$ and, as we
have already discussed in Section 2, the IAM leads 
again to plain ChPT.
That is why we will only use for them data up to 
$\simeq 600$MeV,
although in other channels we are using data at higher energies.

Thus, the values given in Table 3 are
 just a conservative estimate of the range where
we obtain a reasonable fit in the
$(I,J)=(0,0),(2,0),(0,2)$ and $(2,2)$ channels,
when $M_\rho$ fixed to its actual value.
The results are shown in Fig.2, where the
continuous line corresponds to
the central $\hat l_i$ values and the shaded area
to their uncertainties. Notice that the shaded 
area has always been
obtained by varying $\hat l_2$ within its estimated error. 

In Fig.2 it can be seen how it is not possible to fit the $f_0(980)$ 
resonance
with $SU(2)$ ChPT and the IAM. It is clear that, even
though the actual value of the $\delta_{00}$
phase shift may not lie very far from the
unitarized prediction, the qualitative behavior of the curves 
in this channel is not correct above $800$MeV.

With the $\hat l_i$ fit we can obtain the total Breit-Wigner width of the 
$\rho$ resonance from:
\begin{equation}
\Gamma_\rho = \frac{M_\rho^2-s}{M_\rho} \hbox{tan} \delta_{11}(s)
\end{equation}
Indeed we have computed it for different values of $s$ around $M_\rho^2$.
The result is given in Table.3 and it
is quite close to  
the experimental result (see Table 1) 
although slightly high. We will see 
that it is possible to obtain the right 
value when using $SU(3)$ ChPT.

As we have already commented, this result is not at all
trivial, since fitting the right mass does not ensure a correct
description of the resonance. Therefore, even though we are now using the
$M_\rho$ experimental value, the $\Gamma_\rho$ width is again a prediction of
the IAM. In contrast, 
in a unitarization scheme where one introduces
the resonances by hand, one has to
give both the
masses and the widths.
\begin{table}[h]
\begin{center}
\begin{tabular}{|c|c|c|c|c|} \hline
\rule[-3mm]{0mm}{8mm} Method & $\hat l_1$ & $\hat l_2$ &
 $M_\rho$ (input)& $\Gamma_\rho$
\\ \hline\hline
Inverse & & & & \\ 
Amplitude & \raisebox{1.5ex}[1ex]{-0.5 $\pm$ 0.6} &
\raisebox{1.5ex}[1ex]{5.4$\pm$ 0.6} & 
\raisebox{1.5ex}[1ex]{768.8$\pm$ 1.1 MeV}&
\raisebox{1.5ex}[1ex]{155.6$\pm$1.8 MeV} \\ \hline           
\end{tabular}
\end{center}

\leftskip 1cm
\rightskip 1cm
\vskip .2cm

{\footnotesize {\bf Table 3:}
Parameters and results of the one-loop IAM
when $M_\rho$ is fixed to its actual value.}

\leftskip 0.cm
\rightskip 0.cm

\end{table}
\begin{figure}

\vspace{-0.8cm}

\leftskip -2cm
\begin{center}
\mbox{\epsfysize=8.7cm\epsffile{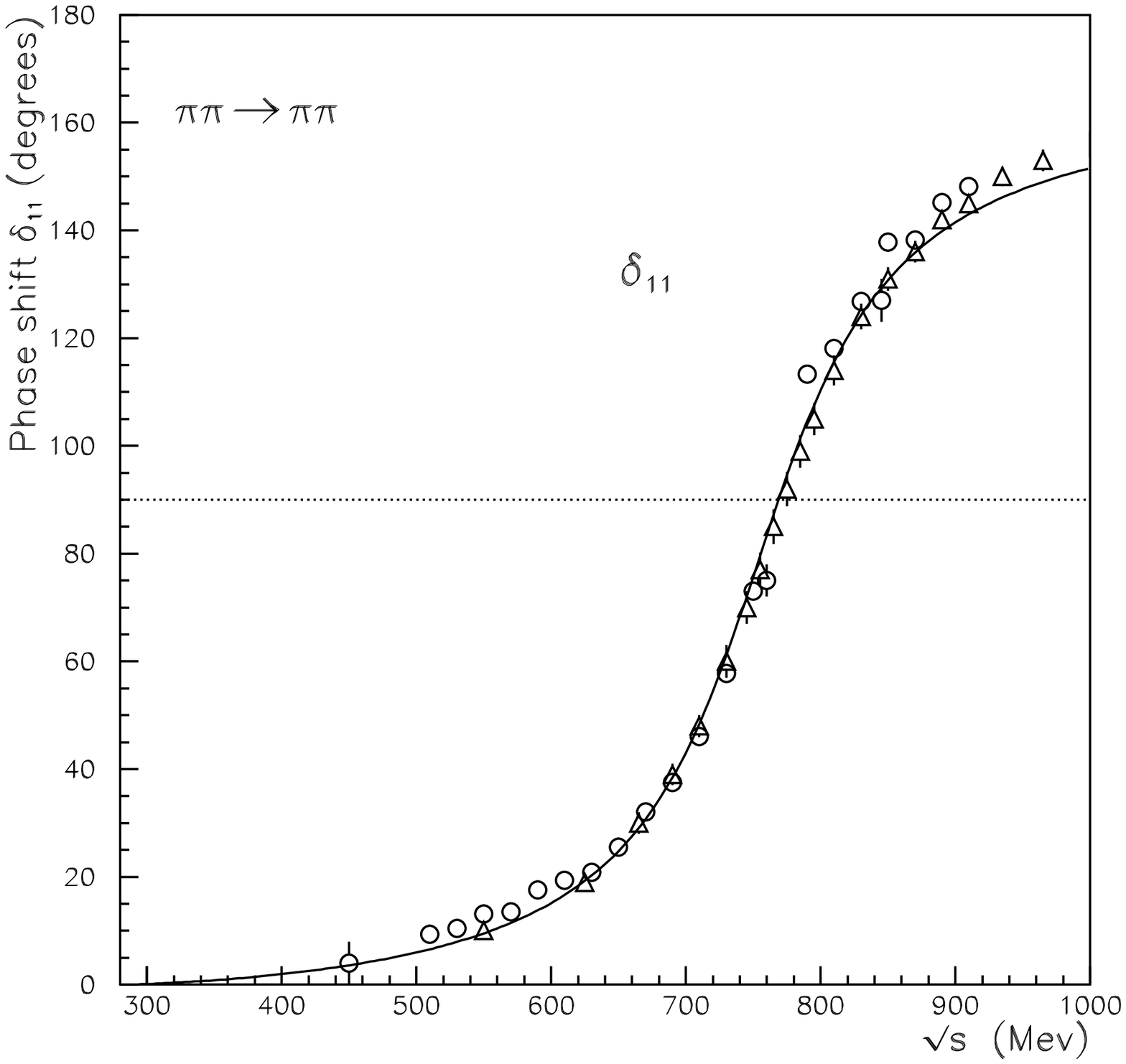}
\epsfysize=8.7cm\epsffile{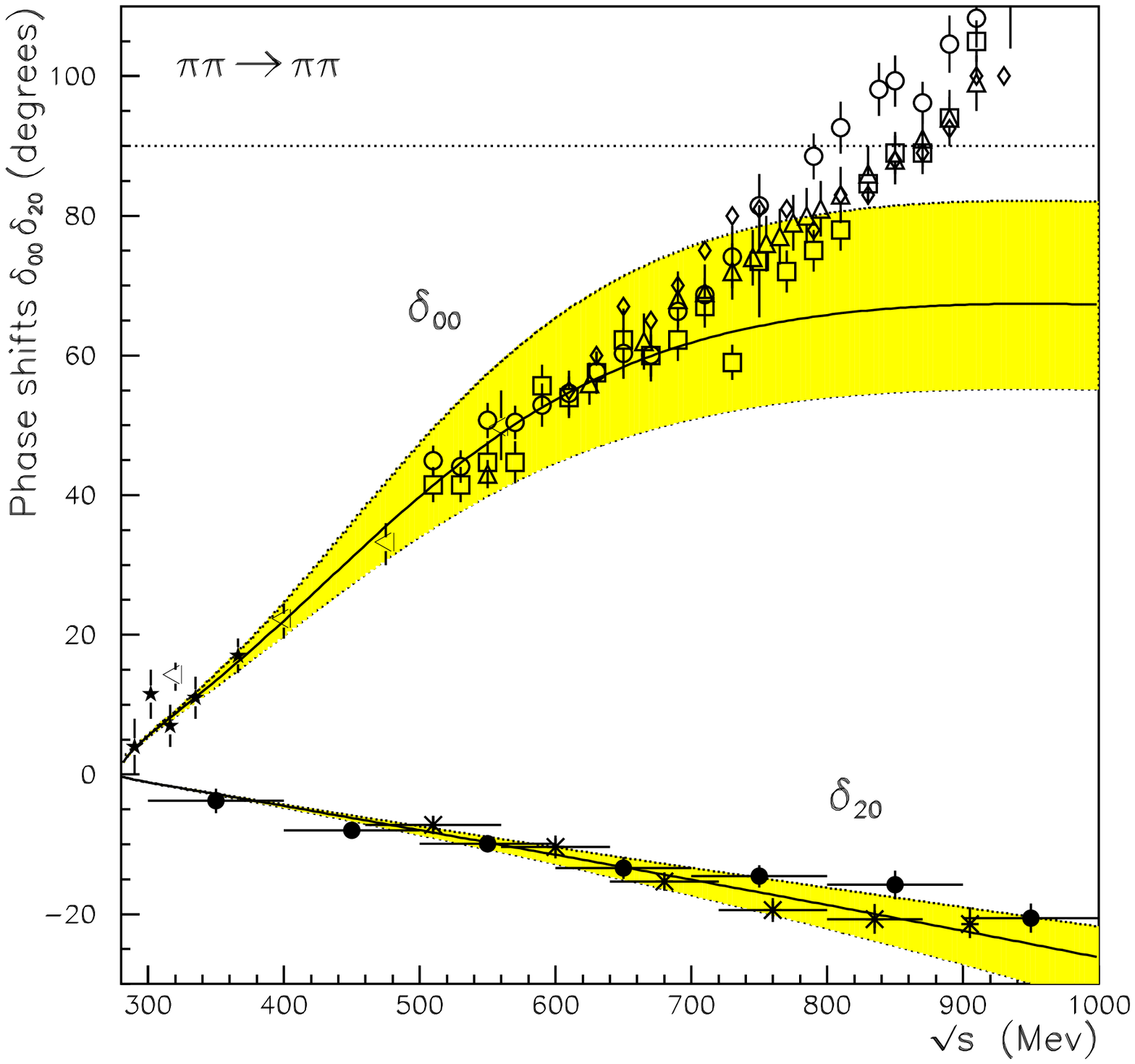}}
\mbox{\epsfysize=8.7cm\epsffile{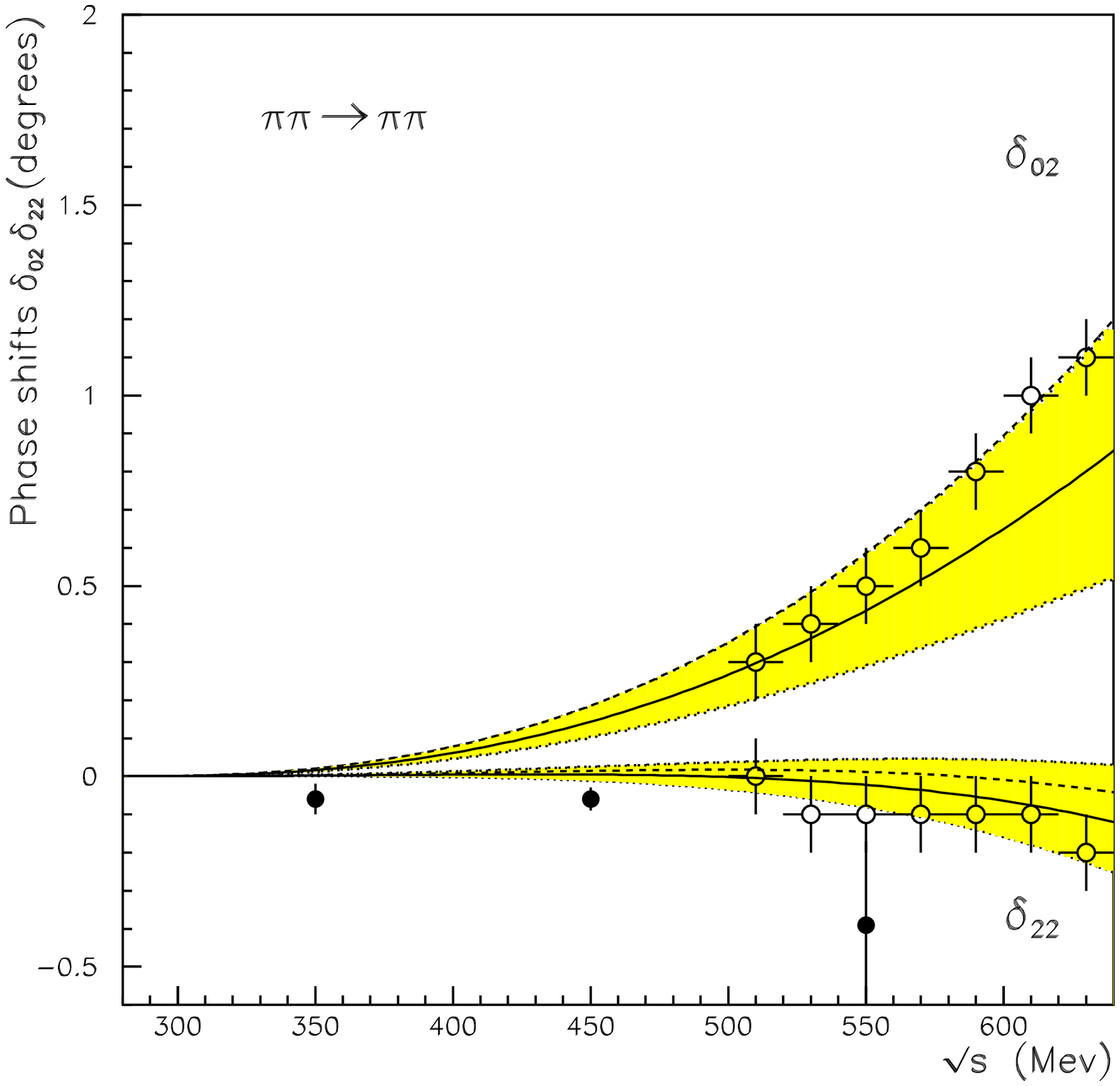}}
\end{center}

\vspace {-.5cm}

\leftskip 1cm
\rightskip 1cm
{\footnotesize {\bf Figure 2.-} Pion elastic scattering phase shifts
$\delta_{IJ}$ obtained from the IAM
fit to the correct $M_\rho$. The shaded areas cover the error bars
of the fitted parameters with the constraint 
$\hat l_1 - \hat l_2=-5.95\pm 0.02$. The dotted straight
lines stand at $\delta=90^0$. Remember that the $J=2$ partial 
waves have to be calculated as in plain ChPT. Indeed, the dashed
lines in those channels correspond to plain ChPT with the 
parameters in the first row of Table 2. 
The symbols for the experimental data
are the same as in Fig.1.
The corresponding curves within $SU(3)$ ChPT would
almost superimpose.}

\leftskip 0.cm
\rightskip 0.cm
\end{figure}

\section{SU(3) Chiral Perturbation Theory}

The extension of the ChPT approach to include the strange quark was done,
once more, by Gasser and Leutwyler \cite{GL2}. In this case there are eight
Goldstone bosons, which are identified with the three pions, the four kaons and
the eta. In principle it is possible to calculate 
the amplitudes of any process
involving any combination of these particles. 
But the thresholds for these
reactions are much higher than in pion scattering, which in practice
restricts severely the effectiveness of the approach. 

Nevertheless, the lowest two particle threshold apart from
$\pi\pi$ scattering is that of
$\pi K$ elastic scattering at $630$MeV, which is 
still within the applicability range of ChPT.
The calculation of this amplitude to $\Od (p^4)$ was
performed by Bernard, Kaiser and 
Mei{\ss}ner \cite{Bernard1,Bernard2} 
who also gave the $\Od (p^4)$ result
for $\pi\pi$ within $SU(3)$ ChPT. 
In the literature, these formulae
 have sometimes appeared with some minor errata which have been corrected in 
the DA$\Phi$NE Physics Handbook \cite{Dhandbook}.
However, even those formulae do not
satisfy perturbative unitarity (see Appendix). 
Following the work in \cite{Bernard1} we have rederived an expression which
does satisfy that requirement, and we have included it
in the Appendix, together with a discussion on how it is obtained
and its  unitarity properties.

In the $SU(3)$ case there are more
 phenomenological parameters 
that we have set to:
\begin{equation}
M_K= 493.65 \mbox{MeV} \;,\; 
M_\eta=548.8 \mbox{MeV} \;,\; 
F_K=1.22 F_\pi \;,\;  F_\eta=1.3 F_\pi
\end{equation}

There are also twelve one-loop parameters, denoted by
$L^r_i(\mu)$. However only 
$L_1^r, L^r_2, L_3, L^r_4, L^r_5, L^r_6$ and $L^r_8$ 
 appear in $\pi K$ in scattering, whereas 
in pion scattering only the following combinations are present:
\begin{eqnarray}
2L^r_1+L_3 &\quad&  L^r_2 \label{combi} \\
\quad 2L^r_4+L^r_5 &\quad& 2L^r_6 + L^r_8 
\end{eqnarray}
Again, and in order to simplify the comparison with previous works,
we have fixed the following values \cite{GL2}:
\begin{equation}
L^r_4(M_\eta)=0\quad L^r_5(M_\eta)=0.0022 \quad
L^r_6(M_\eta)=0 \quad L^r_8(M_\eta)=0.0011
\end{equation}
A precise value of these
parameters is not very important since they are related to the
different masses and decay constants that we had already fixed. Hence, 
in practice,
the only relevant parameters for $\pi\pi$  and $\pi K$ 
scattering in $SU(3)$ are $L_1^r, L_2^r$ and $L_3$.

The IAM was first applied to $SU(3)$ ChPT by the authors
in \cite{UpiK}, were we showed that it reproduces not only the
$\rho(770)$ resonance but also the $K^*(892)$. Our aim in this section is first 
to study the predictive power of the method, whether it can accommodate 
further resonant states, or why it cannot. Then we will present a  
simultaneous fit to $\pi\pi$ and $\pi K$ scattering to the $\rho$ and $K^*$
masses. The new features of this analysis is that it uses the corrected ChPT 
expressions for $\pi K$ scattering which now 
satisfy perturbative unitarity (see the Appendix) and the fact that we also use 
the data on the $J=2$ $\pi\pi$ scattering channels. We will also estimate
 the error bars on the best fit that will be used
to obtain numerical values for some interesting phenomenological
quantities.
This fit will also allow us,
in section 6, to perform a numerical study
of the analytic structure of the IAM amplitudes in
the complex $s$ plane.

\subsection{ Results using low energy parameters}

Let us then start with the IAM
using parameters obtained from 
low energy data. In Table 4 we list different choices of parameters
and methods together with their results for the $\rho$ and $K^*$ 
masses. As in the case of $SU(2)$ ChPT the IAM is able to
predict from low energy data the existence of both resonant states.
Remarkably, the masses thus obtained lie again 
10\% to 15\% away from their actual values.

In Fig.3 we show the result of applying the IAM to $\pi K$ scattering,
with the parameters given in Table 4.
In contrast with plain $\Od(p^4)$ ChPT
(or the K-matrix unitarization method, since they yield
the same phase shifts), it is evident that the IAM
not only accommodates the $K^*$ resonance, but it also 
reproduces the $(3/2,0)$ channel. 

We do not display
the results for $\pi\pi$ scattering in $SU(3)$ because they will almost
superimpose with those in Fig.1. Indeed, the $\bar l_i$ parameters
in lines 2 and 3 of Table 2 were obtained, respectively, from the
$L^r_1,L^r_2,L_3$ in lines 2
and 3 of Table 4 \cite{Rigg,BiGa}, by means of
\begin{eqnarray}  
\bar l_1 &=& 96\pi^2 \left(4L_1^r(M_\eta)+2L_3-\frac{\nu_K}{24}-
\frac{\nu_\pi}{3}
\right )
\nonumber \\
\bar l_2 &=& 48\pi^2 \left( 4L_2^r(M_\eta)-\frac{\nu_K}{12} - \frac{2\nu_\pi}{3} 
\right)
\nonumber \\
\nu_\alpha&=& \frac{1}{32 \pi^2}
\log\left( \frac{M_\alpha^2}{M_\eta^2} \right) \quad; \quad
\alpha= \pi, K
\label{Ll}
\end{eqnarray}

As a matter of fact, we have calculated 
independently the $\pi\pi$ elastic scattering in 
$SU(2)$ and $SU(3)$. Using the above equations 
to relate the parameters in both cases, and below kaon threshold,
 we have obtained the same results up to numerical differences 
($\simeq$1\%),
which would be unobservable in the figures. That is a nice check
of our programs. Therefore, Fig.1 is also the result for $\pi\pi$
 scattering in the $SU(3)$
formalism, but now  with the parameters in Table 4.

\begin{figure}

\vspace{-1.5cm}

\leftskip -2cm
\begin{center}
\mbox{\epsfysize=8.7cm\epsffile{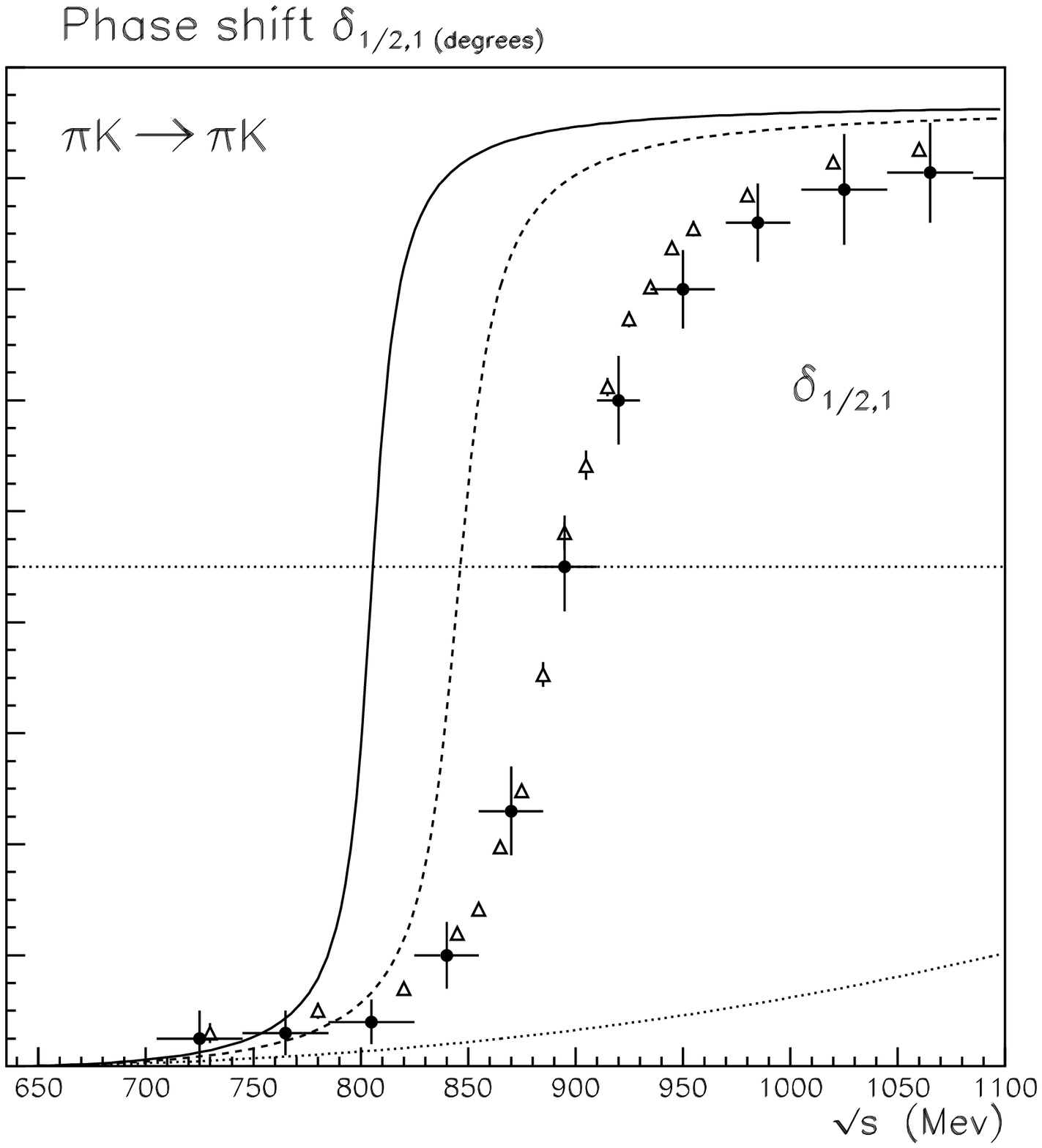}
\epsfysize=8.7cm\epsffile{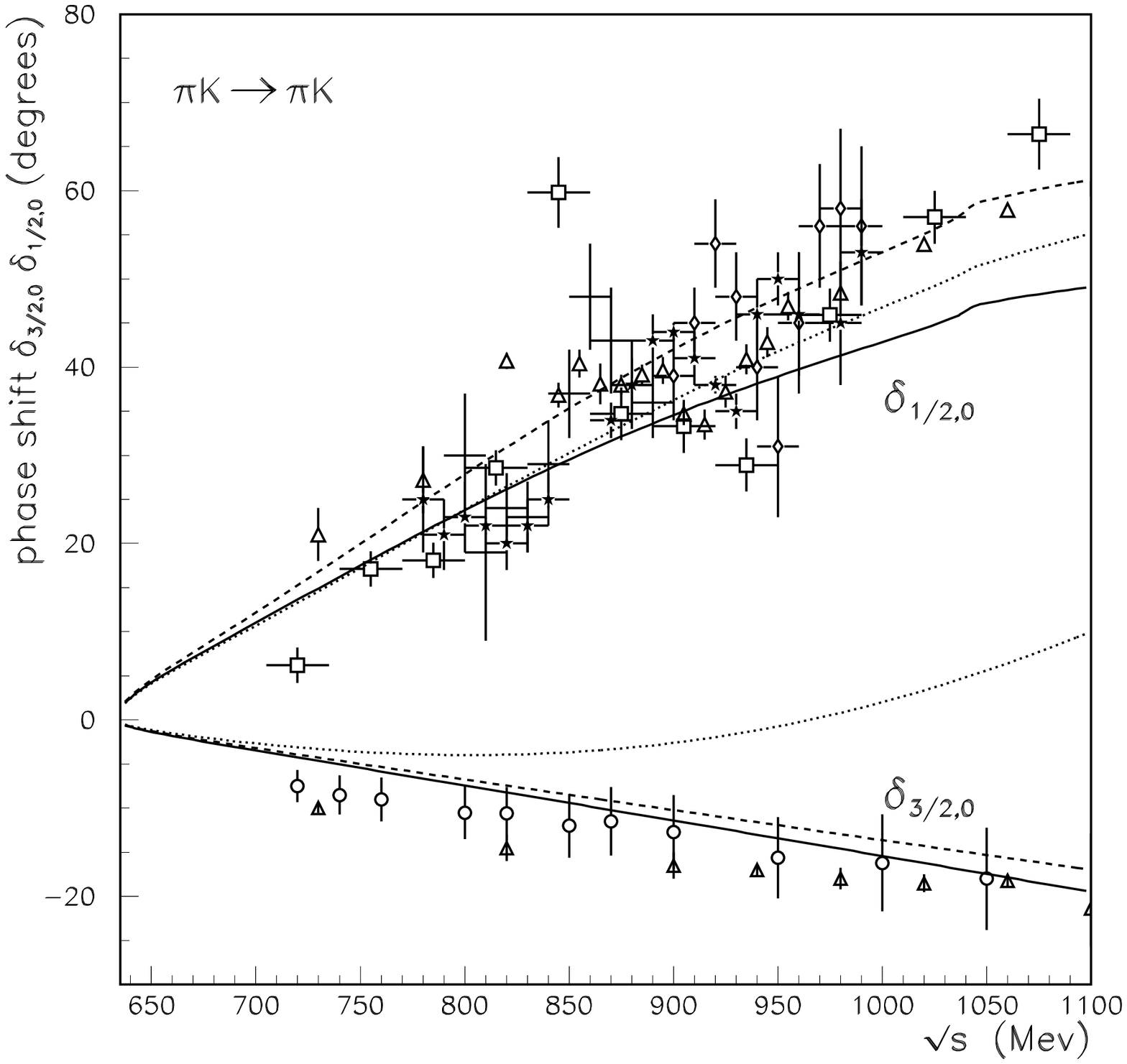}}
\end{center}

\vspace{-0.8cm}

\leftskip 1cm
\rightskip 1cm
{\footnotesize {\bf Figure 3.-} 
Phase shifts for elastic $\pi$K scattering.
The dotted curve is plain ChPT with the $L_i$
parameters in the first line of Table 4. The other two curves
are both obtained from the IAM: the dashed one again with the same parameters
and the continuous one with those in the third line of Table 4.
 The  experimental
data come from:\cite{MeAn71} ($\bullet$), \cite{BiDu72}
($\star$), \cite{LiCh73} ($\circ$), \cite{MaBa74}
($\diamondsuit$), \cite{BaBa75} ($\Box$) and
\cite{EsCa78} ($\triangle$). The straight dotted line
stands at $\delta = 90^0$.}

\leftskip 0.cm
\rightskip 0.cm
\end{figure}

\begin{table}
\begin{center}
\begin{tabular}{|c|c|c|c|c|c|} \hline
\rule[-3mm]{0mm}{8mm}Method & $L^r_1(M_\eta)\cdot10{3}$ & 
$L^r_2(M_\eta)\cdot10{3}$ & $L_3\cdot10{3}$ & $M_\rho$ & $M_{K^*}$ \\ \hline\hline
ChPT & 0.65$\pm$0.28 & 1.89$\pm$0.26 & -3.06$\pm$0.92 & - & - \\ \hline \hline
Inverse & 0.65$\pm$0.28 & 1.89$\pm$0.26 &  -3.06$\pm$0.92 &
 717 MeV & 847 MeV \\ \cline{2-6}
Amplitude & 0.6$\pm$0.3 & 1.75$\pm$0.3 &  -3.5$\pm$1.1 & 680 MeV &
804  MeV\\ \hline
\end{tabular}
\end{center}

\leftskip 1cm
\rightskip 1cm
\vskip .2cm

{\footnotesize {\bf Table 4:}
Different sets of parameters and methods used in the text.
Those of the first two lines come from $K_{l4}$ decays \cite{Rigg}.
Those of the third line come from data on $K_{l4}$ and $\pi\pi$ together with 
some 
unitarization procedure (for details see ref.\cite{BiGa}). The quoted
values of $M_\rho$ and $M_{K^*}$ are calculated with the central values.}

\leftskip 0.cm
\rightskip 0.cm

\end{table}

\subsection{Unitarized fit}

Again we have an expression for
the amplitude that behaves correctly with respect to unitarity and
that presents the right form in the low energy limit. Therefore, we can try to
use the actual $\rho(770)$ 
and $K^*(892)$ masses in order to fit the $\pi\pi$ and $\pi K$
 phase shifts. We remark once more that nothing ensures
that fitting the right masses will give us the right description, 
since, among other things, the widths of the resonances could be wrong.

When dealing with the $SU(3)$ chiral lagrangian we have 
more parameters, and the way they appear in the amplitudes
 is more complicated.
Let us first start with the $\pi\pi$ scattering partial waves in $SU(3)$.
As we have commented in section 3.2.1, in order
to avoid confusions with the ChPT 
low-energy parameters, we will denote the parameters of our fit by $\hat L^r_i$.

 The (1,1) channel only depends on $2L^r_1+L_3-L^r_2$, 
 and will be fixed with $M_\rho$. In so doing we get 
\begin{equation}
2\hat L^r_1+\hat L_3-\hat L^r_2=(-3.11\pm 0.01)10^{-3}
\label{dif}
\end{equation}
As a consistency check we see that it is 
within a 1\% of $-3.14 \  10^{-3}$ which is  
obtained from the $\hat l_i$ parameters 
of the $SU(2)$ case, with the help of Eq.\ref{Ll}. 

Once again we 
use the channels $(I,J)=(0,0),(2,0),(0,2)$ and $(2,2)$ to determine 
the best $\hat L^r_2$ value, which indeed is the same that we would have obtained 
from the $\hat l_2$ $SU(2)$ parameter by means of Eq.\ref{Ll}. It can be found in
Table 5.  Hence, the best $SU(3)$ fit of the $\pi\pi$ phase shifts,
yields almost the same results  as those
obtained with $SU(2)$ and the very same Fig.2 remains valid for
$SU(3)$.
Nevertheless, 
when computing the $\Gamma_\rho$ within the $SU(3)$ formalism, 
we obtain
a much better value than in $SU(2)$, which was about $5$ MeV too high.
It is also listed in Table 5.

Finally we will use $\hat L_3$ to
fix the correct $K^*(892)$ mass.
However, the $K^*(892)$ has an added subtlety, namely, that
the mass splitting between different 
charge states is of the order of 5 MeV. This is 
a small isospin breaking effect that we have not 
included in our approach. Therefore,
we have used an average mass 
$\overline M_{K^*}=894.0\pm2.5$ MeV with an error bar
that includes the mass  of any $K^*(892)$ state, no matter what its charge may be.
That uncertainty has also 
been taken into account in the $\hat L^r_i$ error estimates.

Once we have $\hat L_3$, we use $\hat L_2^r$ and 
Eq.\ref{dif} to obtain $\hat L^r_1$. The parameters of this fit have been 
collected in Table 5, together with 
 $\Gamma_\rho$ and $\Gamma_{K^*}$, which can be considered 
as predictions of the approach. Notice, however, that in this case the
width of the $K^*(892)$ resonance lies 20\% away from its actual value,
which nevertheless is
a reasonably good result in view of the whole
fit in that channel.

\begin{table}
\begin{center}
\begin{tabular}{|c|c|c|c|c|c|} \hline
\rule[-3mm]{0mm}{8mm} Method & $\hat L_1^r(M_\eta) \cdot 10^3$ 
& $\hat L^r_2(M_\eta) \cdot 10^3$ & $\hat L_3 \cdot 10^3$ &  
$\Gamma_\rho$ & $\Gamma_{K^*}$\\ 
\hline\hline
Inverse & & & & & \\ 
Amplitude & \raisebox{1.5ex}[1ex]{0.41$\pm$0.20} &
\raisebox{1.5ex}[1ex]{1.48
$\pm$ 0.33} & \raisebox{1.5ex}[1ex]{-2.44$\pm$ 0.21} & 
\raisebox{1.5ex}[1ex]{149.9$\pm$ 1.2 MeV}&
\raisebox{1.5ex}[1ex]{41.2 $\pm$ 1.9 MeV} \\ \hline           
\end{tabular}
\end{center}

\leftskip 1cm
\rightskip 1cm
\vskip .2cm

{\footnotesize {\bf Table 5:}
Parameters and results of the $SU(3)$ IAM
when $M_\rho=768.8\pm 1.1$MeV and 
$\overline M_{K^*}=894.00\pm 2.5$ MeV are fixed to their actual values.
Notice that for $K^*(892)$ we have chosen an average mass
between its different charge states.} 

\leftskip 0.cm
\rightskip 0.cm

\end{table}

Concerning the $\hat L_i$ parameters, they are
compatible with those in Table 4, which were
obtained from low energy data. Even more, 
they are also consistent with other
parameters obtained from the IAM applied to the form factors
of the $K\rightarrow \pi\pi l \nu$ decays \cite{Hannah}, which
are very well know experimentally:
\begin{equation}
\hat L_1^r(M_\eta)=(0.74\pm 0.14) 10^{-3} \quad
\hat L_2^r(M_\eta)=(1.07\pm 0.18) 10^{-3} \quad
\hat L_3(M_\eta)=(-2.45\pm 0.52) 10^{-3} 
\end{equation}
(notice that in that reference they are using 
$F_K=F_\pi$, so that the parameters do 
necessarily differ).

Nevertheless, it would not make any sense to try to reduce
the error bars of these parameters.
We consider that the approach that we have been following
here can only be consistent within a few percent error
level. In order to have a better accuracy
it would be necessary to take into account higher
order ChPT corrections, isospin-breaking effects 
and the whole approach should be modified 
following the comments that we made in previous sections.

In Fig.4 we show the results of the $SU(3)$ IAM fit to the resonance masses,
in terms of elastic scattering phase shifts, which we 
think deserve some comments:
\begin{itemize}

\item First notice that we are not showing the curves 
for $\pi\pi$ scattering because they are exactly those in Fig.2. 
The differences only appear above the two kaon threshold, since
in the $SU(3)$ formulae we are also considering internal loops
of kaons and etas.

\item In the $\pi K \rightarrow \pi K$ case we can extend  
the graphs up to 1100 MeV, or even more.
 The reason is that the first two body inelastic threshold is
$K \eta$ production at 1040 MeV and, in contrast to the $\pi\pi$ case, there 
is no nearby resonance. Indeed, the next resonant state in $\pi K$ elastic 
scattering is $K_0^*(1430)$, very high to affect dramatically our results 
at $1100$ MeV, but also to be correctly reproduced by the IAM method.
Nevertheless, the existence of the $K \eta$ threshold can be noticed in
the $I=1/2,J=0$ channel, as a small bump in the curves at precisely
1040 MeV.

\item The shaded area in the $K^*(892)$ channel is not only due to
the averaged mass for $K^*(892)$ with $2.5$MeV error, but also to
the fact that we have to
determine several parameters to get the right mass, 
in contrast with the $\rho(770)$ case, when we only had to fix one.

\end{itemize}

\begin{figure}

\vspace{-0.8cm}

\leftskip -2cm
\begin{center}
\mbox{\epsfysize=8.7cm\epsffile{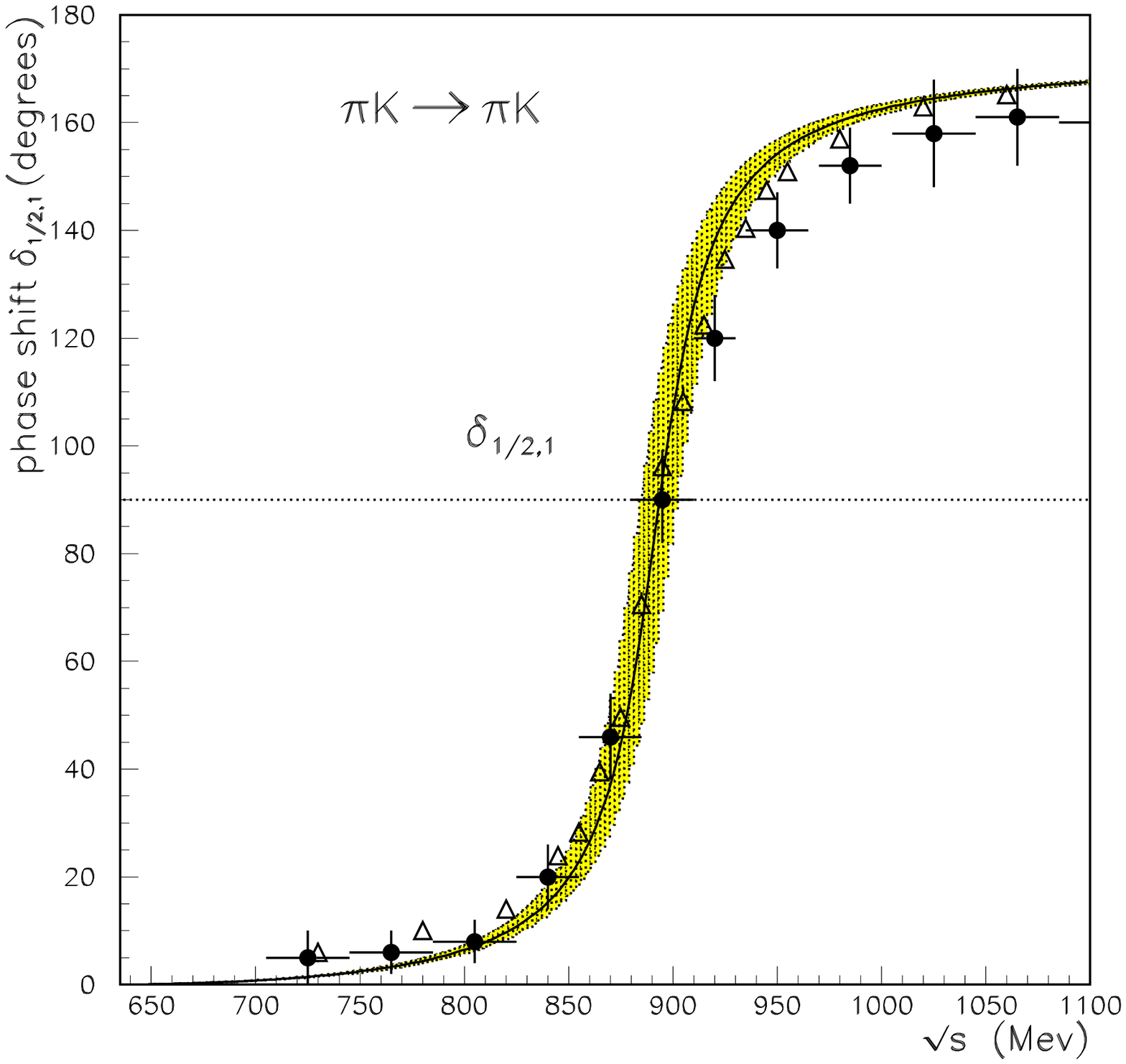}
\epsfysize=8.7cm\epsffile{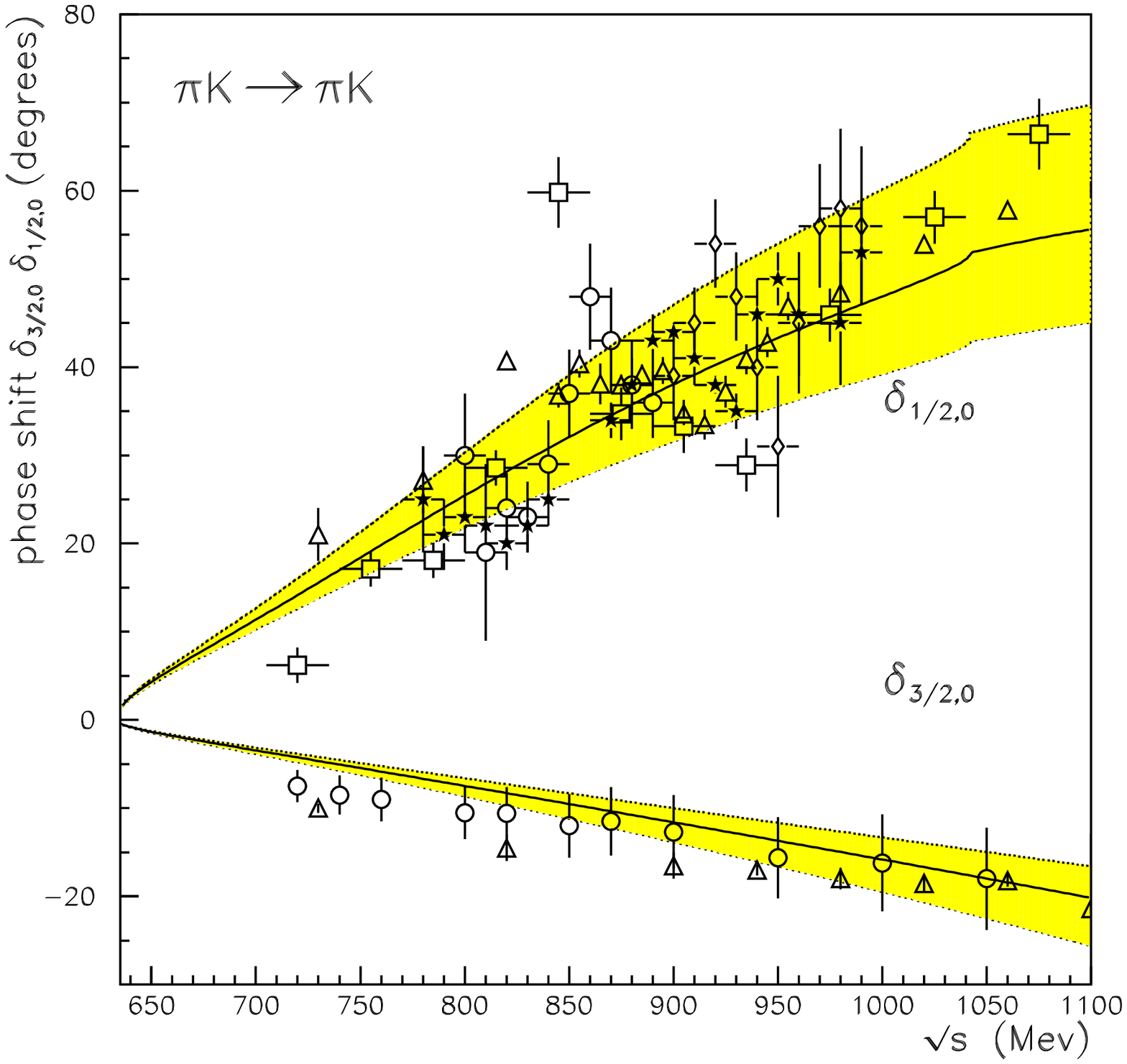}}
\end{center}

\vspace {-.5cm}

\leftskip 1cm
\rightskip 1cm
{\footnotesize {\bf Figure 4.-} $\pi K$ elastic scattering phase shifts
$\delta_{IJ}$ obtained from the IAM
fit to the correct $M_\rho$ and $M_{K^*}$. 
The shaded areas cover the error bars
of the fitted parameters with the constraint 
$2\hat L^r_1+\hat L_3-\hat L^r_2=(-3.11\pm 0.01)10^{-3}$. 
The dotted straight
line stands at $\delta=90^0$. 
The symbols for the experimental data
are the same as in Fig.3.}

\leftskip 0.cm
\rightskip 0.cm
\end{figure}
We have explicitly checked that our ChPT amplitudes 
satisfy perturbative unitarity. As it is explained in the Appendix, 
previous calculations \cite{Bernard2,UpiK}, 
including ours, did not 
respect this condition, although by a very small amount.
That is why the values of the best parameters for
this fit are slightly different from those of our previous 
work \cite{UpiK}.

\vskip .5cm

{\bf Phenomenological parameters}

\vskip .5cm

Once we have a good parametrization of $\pi\pi$ and
$\pi K$ elastic amplitudes, we can use it to obtain 
the values of some relevant phenomenological parameters.
First we can calculate the scattering lengths, which determine the 
strength of the interactions at low energy.
Despite our IAM fit makes use of high energy data, we
expect that it will reproduce the low energy behavior since
in the low energy limit it reduces to the chiral expansion,
which at ${\cal O}(p^4)$ already yields quite good 
values (see Tables 6 and 7).
However, as far as the IAM is non-perturbative
we are also taking into account higher order effects, that will
modify the results. Indeed, some of these
lengths have already been calculated with the IAM 
and it yields slightly better results than plain ChPT \cite{Hannah}.
We have made again the calculation with our fit, but
as far as we 
have an estimate of the error bars in the $\hat L_i$ parameters, 
we will also give the error estimates 
coming {\em only} from the uncertainties in $\hat L_i$
(mostly dominated by that of $\hat L_2$).

Before giving the results, it is convenient to recall that
the scattering lengths have two different
normalizations. Namely:
\begin{equation}
\Rea t_{IJ}(s)= q^{2J}\left( a_J^I + b_J^I q^2 + \Od(q^4) \right)
\end{equation}
for $\pi\pi$ scattering, where $q$ is the C.M. momentum $q^2=s/4-M_\pi^2$,
and
\begin{equation}
\Rea t_{IJ}(s)= \frac{\sqrt{s}}{2}
q^{2J}\left( a_J^I + b_J^I q^2 + \Od(q^4) \right)
\end{equation}
for $\pi K$ scattering, where now 
$q^2=[s-(M_K+M_\pi)^2][s-(M_K-M_\pi)^2]/4s$.

The predictions of our fit for the $\pi\pi$ and $\pi K$ 
scattering lengths are given in Tables 6 and 7
(in $M_\pi$ units). Notice that
all the values are compatible with the experimental data,
and in general they only differ very slightly from the $\Od(p^4)$
ChPT results, usually in the right direction toward the central 
value. However, the experimental error bars are still too big to
arrive at any conclusion. Also the error bars in the IAM have to be
interpreted cautiously, since they are obtained {\em only} from the
uncertainties in the $\hat L_i$ parameters.

As we have already commented, very recently there has appeared a two
loop calculation of $\pi\pi$ scattering within SU(2) ChPT.
It estimates $a^0_0\sim 0.217$ or $0.215$
and $a^0_0-a^2_0\sim 0.258$ or $0.256$, which are precisely the
values obtained with our IAM fit. This fact gives support to 
the idea that the IAM 
somehow takes into account higher order terms even at low energies.

Notice that we do not compare with the two-loop calculation
of the scattering lengths and slopes in \cite{Knecht} 
because they have used them as an {\em input} in a $\chi^2$-fit
to determine their additional $\alpha$ and $\beta$ parameters.
Therefore, their values are almost exactly those of 
the experimental data. However, as far as there is no data 
for the $b^1_1$ $\pi\pi$ slope parameter, their value can be 
regarded as a prediction. They give $b^1_1=(0.54\pm0.15)10^{-2}$,
which is consistent with our result and with $b_1^1=(0.6\pm0.4)10^{-2}$,
that was obtained from sum rules in \cite{Toublan}.

We have also calculated the phase of the $\epsilon'$ parameter, which
measures direct CP violation in $K\rightarrow\pi\pi$ decays 
\cite{Wu}. It is related to the $s$-wave phase shifts as follows:
\begin{equation}
\phi(\epsilon')= 90^o-(\delta^0_0-\delta^2_0)_{s=M_{K^0}^2}
\end{equation}
Our result is:
\begin{equation}
\phi(\epsilon')= (42^{-7}_{+5})^o
\end{equation}
very close to $\phi(\epsilon')=(45\pm 6)^o $ which is
 obtained in plain ChPT \cite{GaMei}. In contrast with the case
of the scattering lengths, the value of this angle is not used
as an input in \cite{Knecht} and is therefore a prediction 
of their best fit. The value they quote is
$\phi(\epsilon')=(43.5\pm 2 \pm 6)^o $.

\begin{table}
\begin{center}
\begin{tabular}{lccc}
\rule[.9cm]{0mm}{1.5mm}
\raisebox{1.5mm}{$a^I_J$ } &
\raisebox{1.5mm}{ChPT} &
\raisebox{1.5mm}{IAM fit} &
\raisebox{1.5mm}{Experiment} \\ \hline \hline
$a^0_0$ & 0.201 &
0.216 $\pm$ 0.008 &
0.26 $\pm$ 0.05\\
$b^0_0$ & 0.26 &
0.289 $\pm$ 0.025 &
0.25 $\pm$ 0.03\\ \hline
$a^2_0$ & -0.041 &
-0.0417 $\pm$ 0.0014 &
-0.028 $\pm$ 0.012 \\
$b^2_0$ & -0.070 &
-0.075 $\pm$ 0.003 &
-0.082 $\pm$ 0.008\\ \hline
$a^1_1$ & 3.6 $\cdot$ 10$^{-2}$ &
(3.744 $\pm$ 0.002)$\cdot$10$^{-2}$ &
(3.8 $\pm$ 0.2)$\cdot$10$^{-2}$\\
$b^1_1$ & 0.43$\cdot$10$^{-2}$ &
(0.515 $\pm$ 0.001)$\cdot$10$^{-2}$ &
-  \\
\hline
$a^0_2$ & 20 $\cdot$ 10$^{-4}$ &
(17.1 $\pm$ 3.5)$\cdot$10$^{-4}$ &
(17 $\pm$ 3)$\cdot$10$^{-4}$\\
\hline
$a^2_2$ & 3.5 $\cdot$ 10$^{-4}$ &
(2.8 $\pm$ 1.5)$\cdot$10$^{-4}$ &
(1.3 $\pm$ 3.1)$\cdot$10$^{-4}$ \\
\hline \hline
\end{tabular}

\end{center}

\leftskip 1cm
\rightskip 1cm
\vskip .1cm

{\footnotesize {\bf Table 6:}
$\pi\pi$ scattering lengths. The one-loop ChPT results
are taken from \cite{Rigg}. The experimental data
come from \cite{Petersen}. The errors in the IAM fit
come only from the uncertainties in the parameters . 
They do not include other theoretical uncertainties.}

\leftskip 0.cm
\rightskip 0.cm
\end{table}
\begin{table}
\begin{center}
\begin{tabular}{lccc}
\rule[.9cm]{0mm}{1.5mm}
\raisebox{1.5mm}{$a^I_J$ } &
\raisebox{1.5mm}{ChPT} &
\raisebox{1.5mm}{IAM fit} &
\raisebox{1.5mm}{Experiment} \\ \hline \hline
$a^{3/2}_0$ & -0.043 &
-0.049 $\pm$ 0.004 &
-0.13...-0.05\\
$b^{3/2}_0$ & - &
-0.026 $\pm$ 0.003 &
- \\ \hline
$a^{1/2}_0$ & 0.148  &
0.155 $\pm$ 0.012 &
0.13...0.24 \\
$b^{1/2}_0$ & - &
0.087 $\pm$ 0.016 &
-\\ \hline
$a^{1/2}_1$ & 0.012 &
0.0146 $\pm$ 0.0012 &
0.017...0.018\\
\hline \hline
\end{tabular}

\end{center}

\leftskip 1cm
\rightskip 1cm
\vskip .1cm

{\footnotesize {\bf Table 7:}
$\pi K$ scattering lengths. Note that the ChPT results
have been obtained using the corrected
formulae in the Appendix. The experimental data
come from \cite{Bernard1}}

\leftskip 0.cm
\rightskip 0.cm
\end{table}
Finally, in Fig.5 we show the phase difference $\delta_{00}-\delta_{11}$,
compared with the hitherto 
available experimental data \cite{Rosselet}. The difference between
the IAM and plain ChPT at high energies is due to the presence of
the $\rho$ resonance. Nevertheless, there are also some differences at
low energies, since the dispersive approach is somehow taking into account
higher order contributions. 

\begin{figure}

\vspace{-0.8cm}

\leftskip -2cm
\begin{center}
\mbox{\epsfysize=8.7cm\epsffile{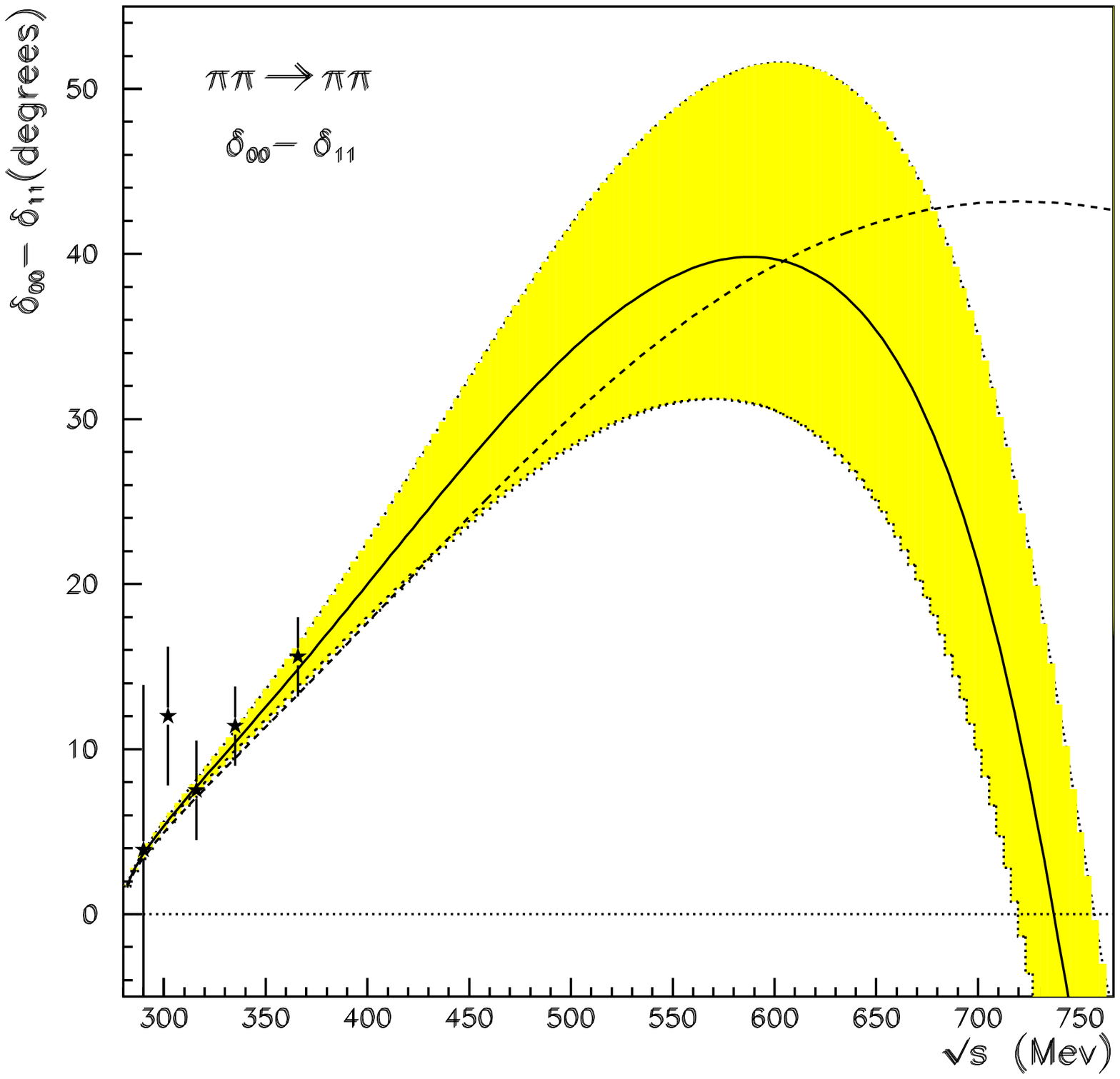}}
\end{center}

\vspace {-.5cm}

\leftskip 1cm
\rightskip 1cm
{\footnotesize {\bf Figure 5.-}
$\delta_{00}-\delta_{11}$ phase shift difference from the IAM fit
(solid line) and plain ChPT (dashed line). The shaded area
covers the uncertainty in the $\hat L_i$ parameters and
the data come from \cite{Rosselet}. }

\leftskip 0.cm
\rightskip 0.cm
\end{figure}

\section{The IAM in the complex $s$ plane}

The main objection to unitarization procedures is the
apparent arbitrariness in their predictions, which may differ from 
one another. In most cases, these methods are nothing but a
small modification of the amplitudes
so that they can satisfy the
unitarity constraint in Eq.\ref{uni}, while keeping
at the same time the good low energy behavior. 
But that constraint is not enough to determine the
amplitude completely.
Thus there are as many unitarization techniques as
algebraic tricks to implement such a constraint exactly or
to get a better approximation. 

However, we have 
already seen in Section 3.1 that, below any other inelastic threshold,
the inverse amplitude method can be derived directly from the
analytic structure of the general two body elastic scattering amplitude.
Our purpose in this section is to show that, 
apart from satisfying elastic unitarity,
it provides the correct
analytic structure required from relativistic Quantum Field Theory.
Such an structure is not trivial at all and cannot be
reproduced by other unitarization procedures.
Both the left and right unitarity cuts
are already present in plain ChPT, therefore, we will mainly focus
on the poles in the second Riemann sheet.

In the previous section we used the most naive criteria to identify
resonances, i.e., that the phase crosses the $\delta=90^0$ value.
However, that is only true for the simplest cases.
The rigorous characterization of resonances is made in terms
of poles in the second Riemann sheet of the amplitudes
in the $s$ complex plane.
Indeed, when a resonance is produced by just one of these poles,
both its mass and width are related to the pole position by
\begin{equation}
\sqrt{s}_{pole} \simeq M_R + i \frac{\Gamma_R}{2} 
\label{respar}
\end{equation}
provided the width is small enough.

In this work we have extended to the $s$ complex plane 
both the $\pi\pi$ and the $\pi K$ elastic scattering IAM amplitudes
obtained in the previous section.
Notice that the cuts in ChPT come from logarithmic functions, so that
we have infinite sheets in the complex plane. However, only two
of them correspond to the first and second Riemann sheets.
Once we have identified these sheets
we can check whether the resonances that
we found in previous sections
are produced by a pole in the second Riemann sheet
and thus whether they have a real sound basis.

\begin{figure}

\vspace{-0.8cm}

\leftskip -2cm
\begin{center}
\mbox{\epsfysize=6.5cm\epsffile{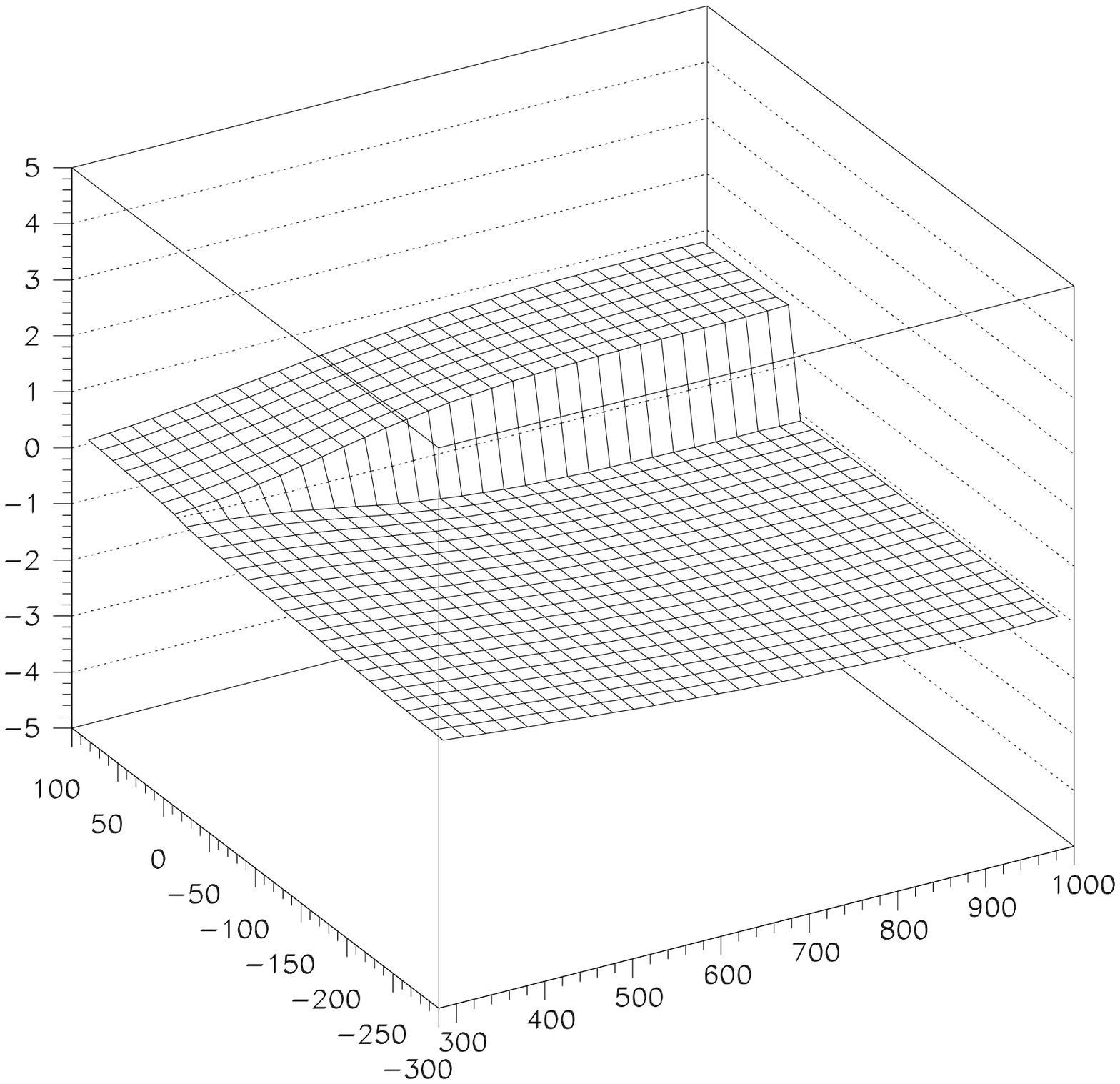}
\epsfysize=6.5cm\epsffile{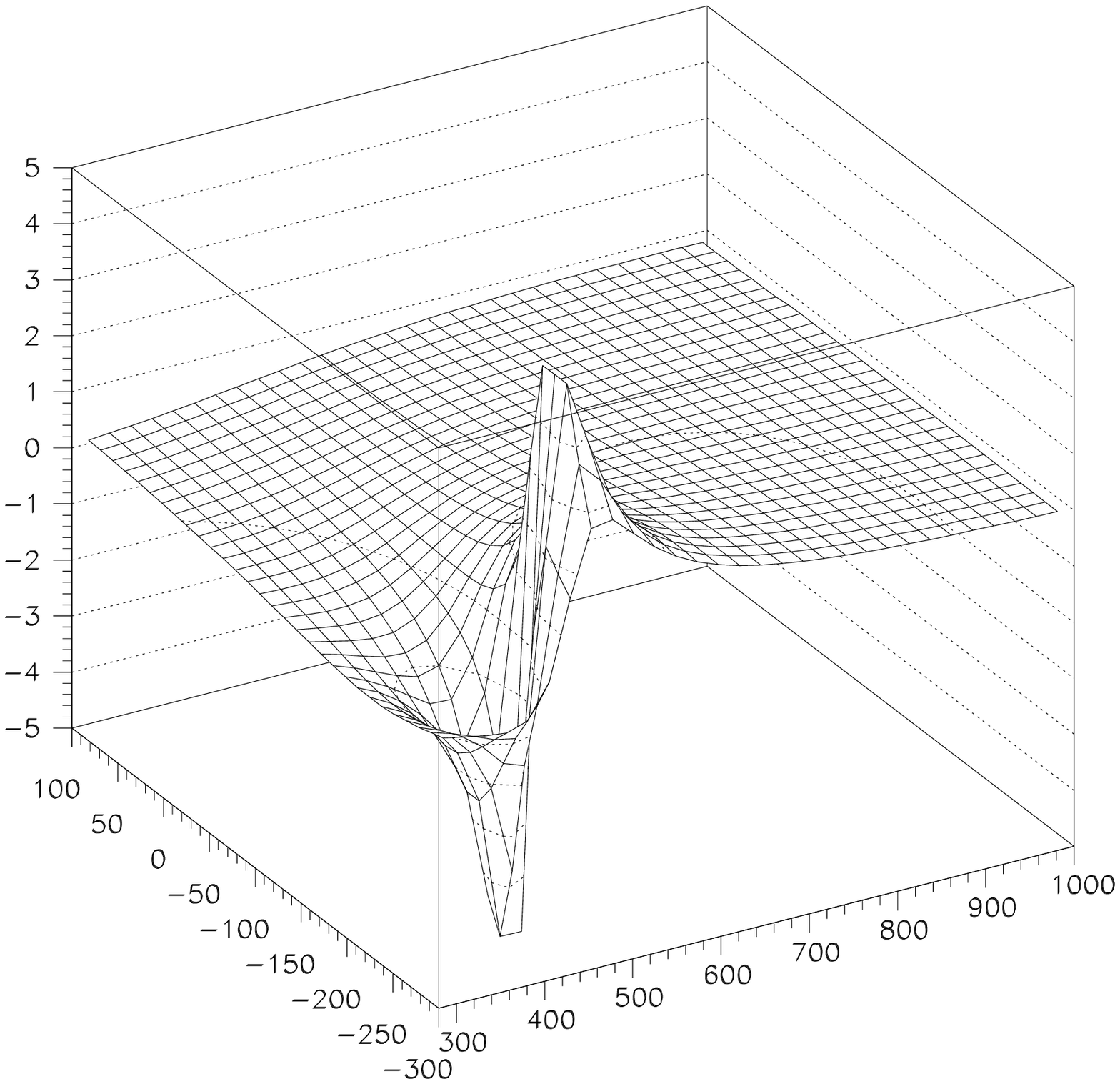}}
\mbox{\epsfysize=6.5cm\epsffile{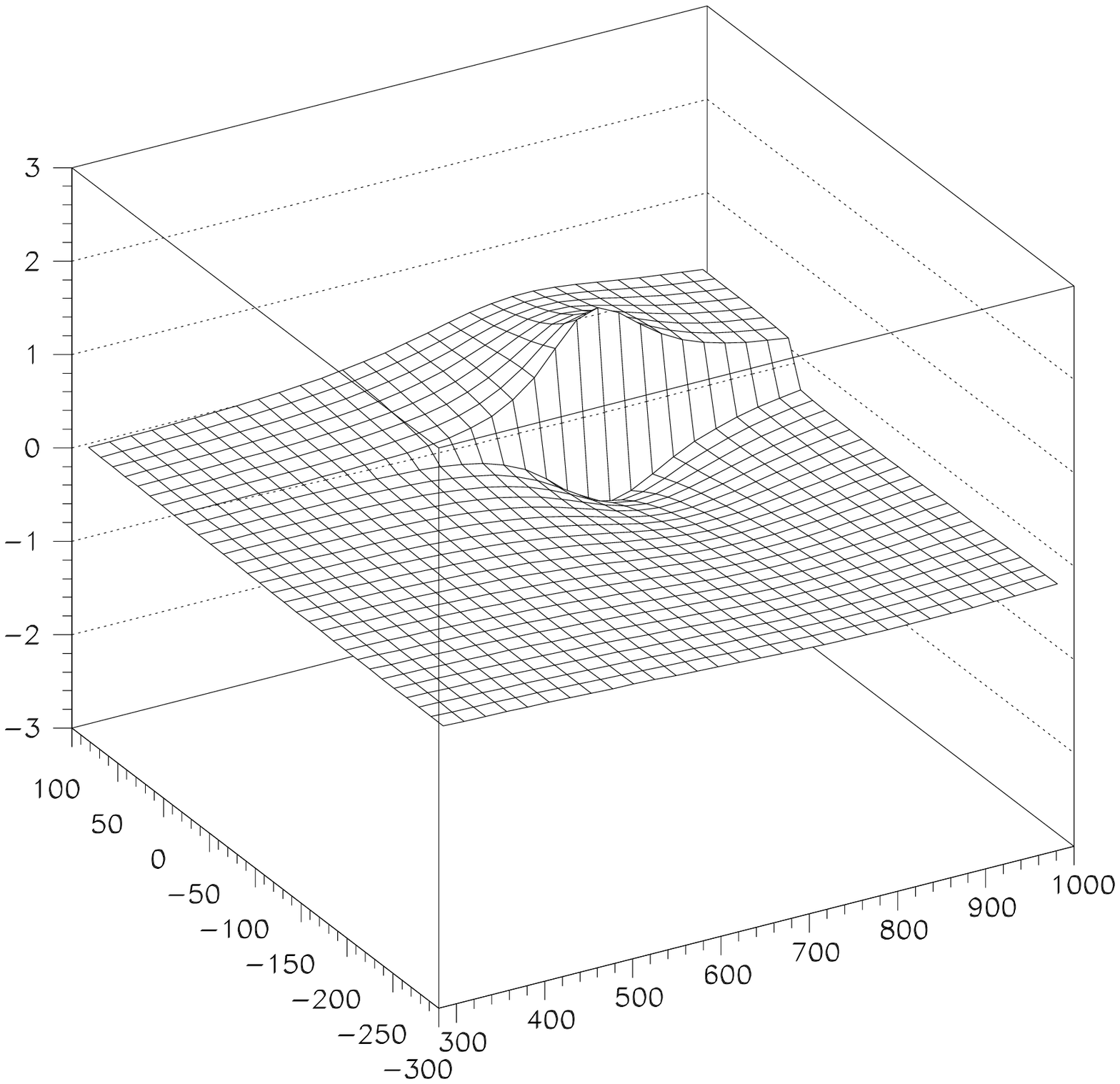}
\epsfysize=6.5cm\epsffile{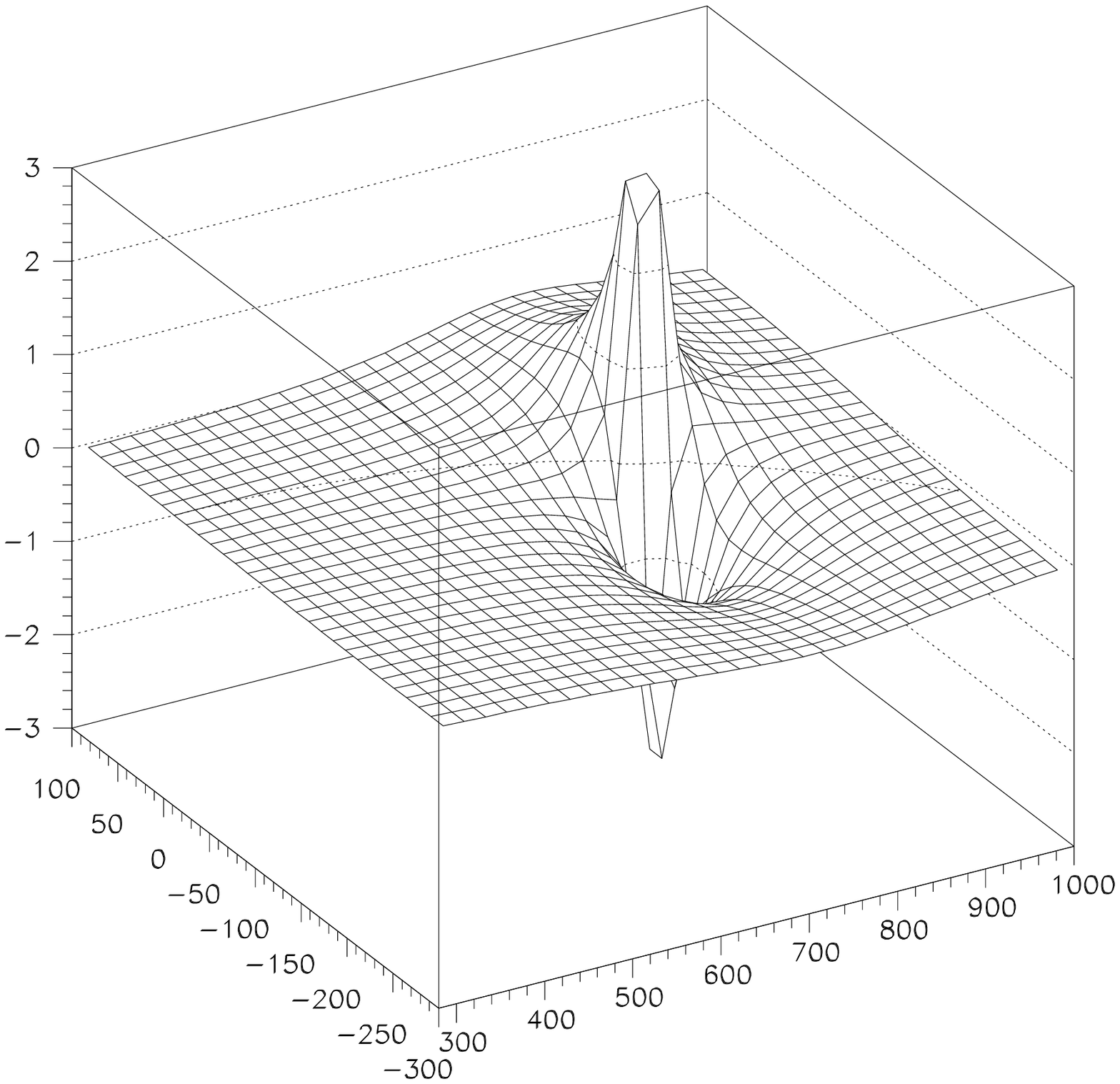}}
\mbox{\epsfysize=6.5cm\epsffile{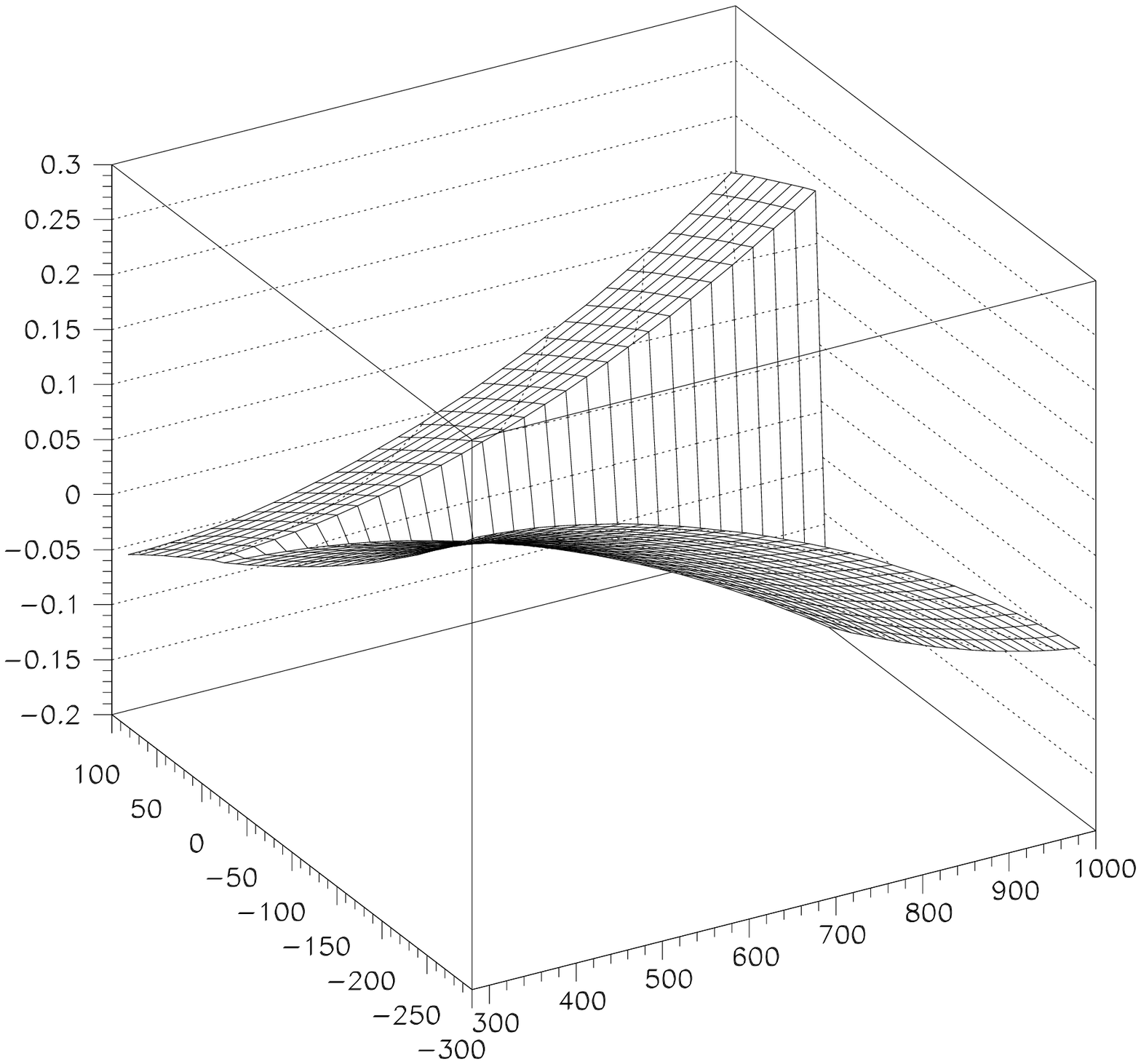}
\epsfysize=6.5cm\epsffile{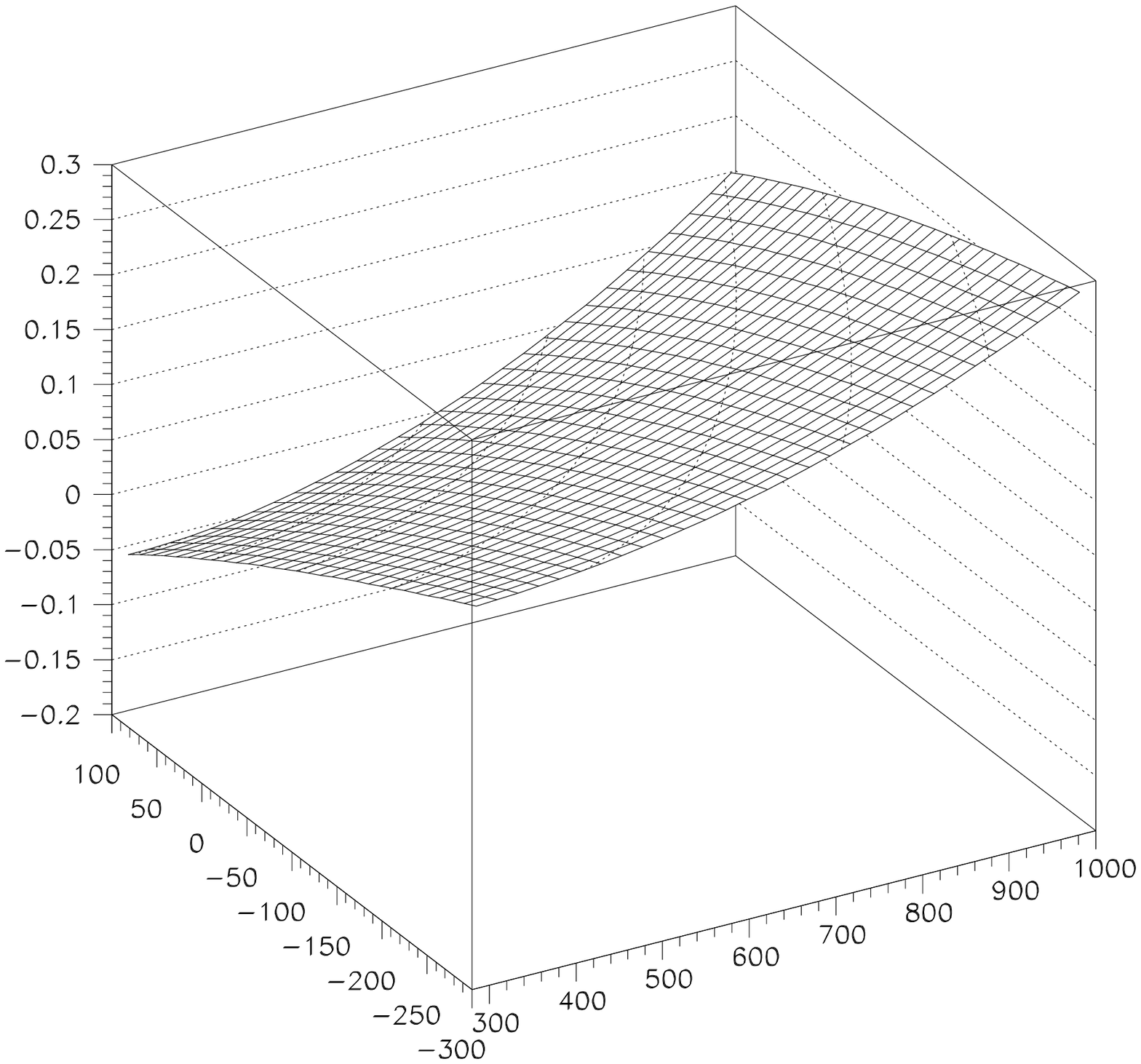}}
\end{center}

\vspace {-.5cm}

\leftskip 1cm
\rightskip 1cm
{\footnotesize {\bf Figure 6.-} Imaginary parts of
the $\pi\pi \rightarrow \pi\pi$ amplitudes in the complex $s$ plane.
The first row is the $(I,J)=(0,0)$ channel, the second is $(1,1)$
and the bottom is $(2,0)$. The left plots correspond to the first
Riemann sheet, and those on the right, to the second.}

\leftskip 0.cm
\rightskip 0.cm
\end{figure}

We will first analyze the $\pi\pi \rightarrow \pi\pi$ process.
In Figure 6 we represent the imaginary part of the amplitude in the
complex $s$ plane for the three channels $(I,J)=(0,0),(1,1)$ and $(2,0)$.
Notice that when we say complex $s$ plane, we mean that we have 
parametrized $s$ as $s=(E+iC)^2$, where $E$ is the CM energy and is
represented in the real axis whereas $C$ provides the complex part.
On the left column we have displayed the results in the first Riemann sheet,
whereas in the right column
we have continued through the cut to the 
lower half of the second Riemann sheet. In all cases it can be 
clearly noticed the existence of a cut on the real axis on the first 
Riemann sheet. As we had commented before, a right cut is not anything
completely new, since it is already present
in one-loop ChPT, although in that case, the values that the amplitudes
take on it are different. In contrast, the most striking new feature 
in the IAM amplitudes
is the appearance of poles in the second Riemann sheet and how
they determine the amplitude shape for the physical values of $s$.

Indeed, we have found two poles with $\Ima s<0$ in the second Riemann sheet,
one in the $(0,0)$ partial wave and another one in $(1,1)$. Let us start
with the second, which clearly corresponds to the $\rho$ resonance.
The position of this pole can be obtained from the contour plots
in Fig.7, and it is found at around $E_R\sim 760 -i 75$.
Using Eq.\ref{respar} we see that it is in a good agreement with the
$\rho(770)$ mass and width parameters given in Table 1. Therefore,
we can conclude that this pole is completely consistent
with the $\rho(770)$ resonance. 

The other pole that can be seen in $\pi\pi \rightarrow \pi\pi$ is
on the $(0,0)$ channel. Using the parameters of the best $SU(3)$ IAM
fit of the previous section, we find that it is located at 
$E_R\sim 440-i 245$. It is not responsible for
the appearance of any resonance, since it is very far away from the real 
axis. However, from purely phenomenological fits to pion scattering data
it had already been pointed out the existence of such a 
pole around $E_R\sim 408 - i 342$
MeV \cite{Zou}. This pole is responsible for the strong interaction in
that dominates the at low energy the two pion $(0,0)$ channel.
We can now see that even in the channel where there is not an apparent improvement,
the IAM yields the correct analytic structure.

\begin{figure}

\vspace{-0.8cm}

\leftskip -2cm
\begin{center}
\mbox{\epsfysize=5.6cm\epsffile{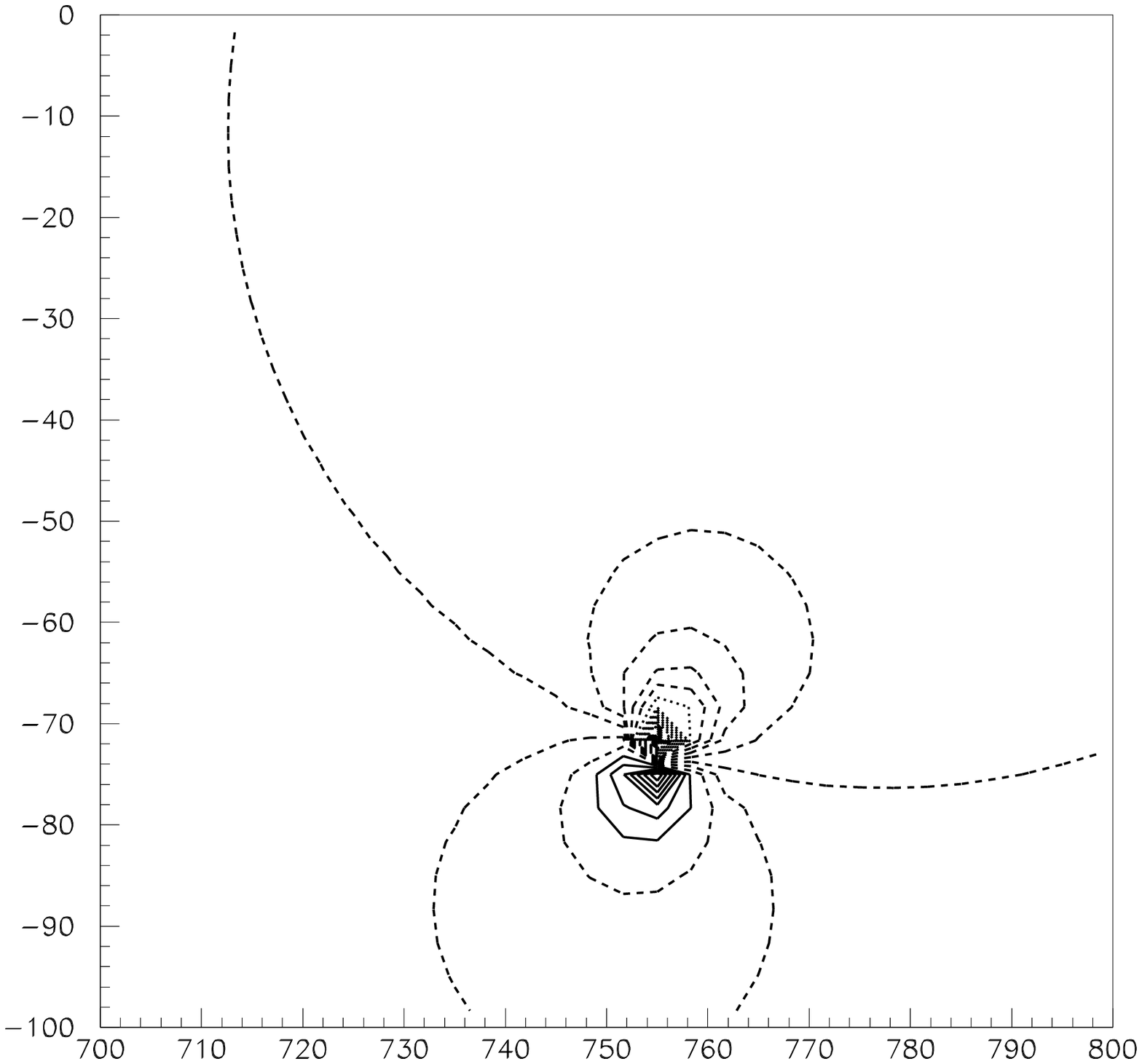}
\epsfysize=5.6cm\epsffile{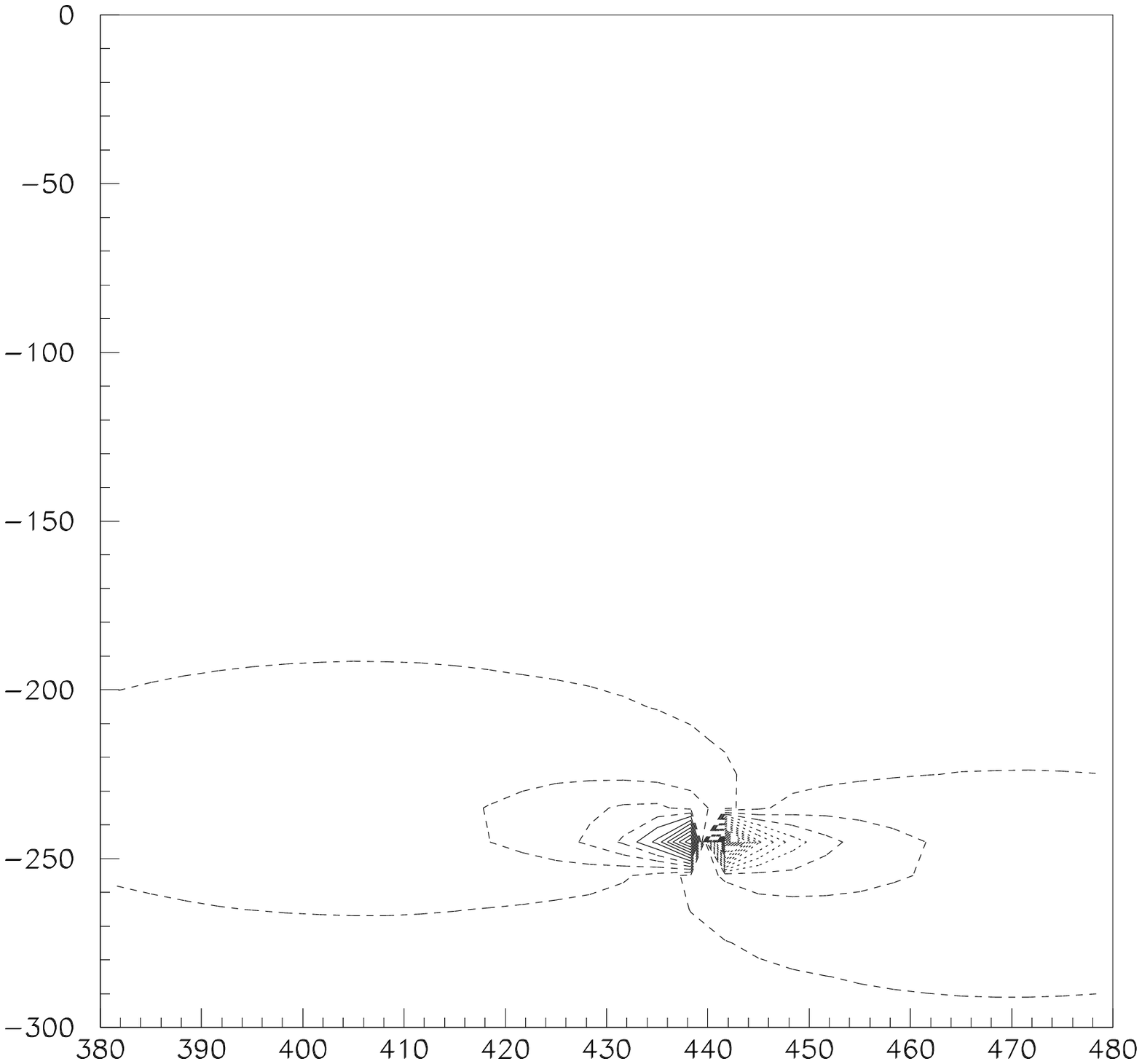}
\epsfysize=5.6cm\epsffile{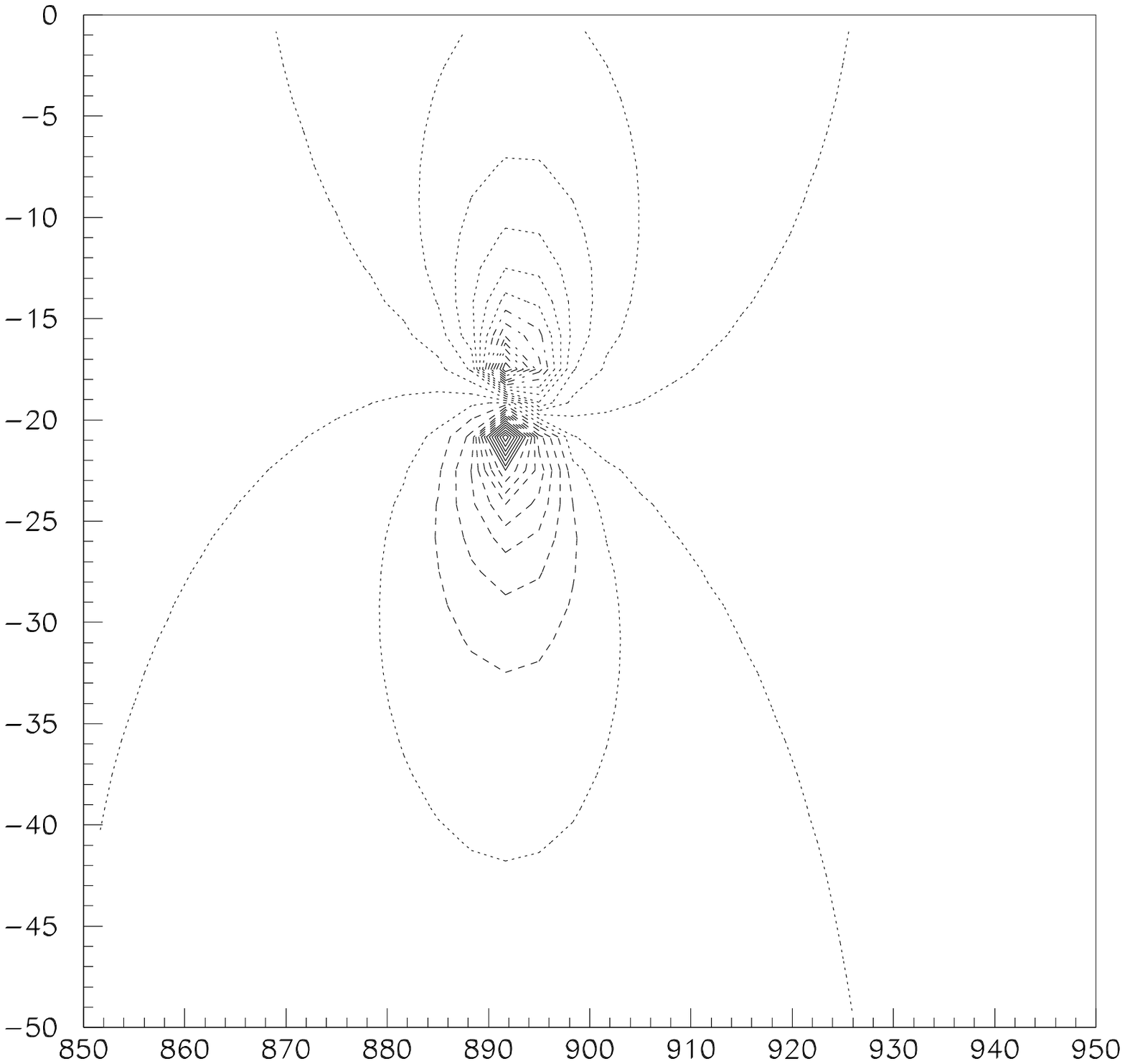}}
\end{center}

\vspace {-.5cm}

\leftskip 1cm
\rightskip 1cm
{\footnotesize {\bf Figure 7.-} Contour plots
of the second Riemann sheets for different $SU(3)$ ChPT unitarized
amplitudes.
From left to right they correspond to the $(1,1)$ and $(0,0)$
$\pi\pi$ scattering channels and the $(1/2,1)$ $\pi K \rightarrow \pi K$
 channel.}

\leftskip 0.cm
\rightskip 0.cm
\end{figure}

Much as it happened in previous sections, the method is not able to 
reproduce the $f_0(980)$ resonance. As we already commented, the
interpretation in terms of poles of this resonance is still controversial.
Following the same steps as before, we have also identified 
the four Riemann sheets that now appear due to the superposition of two
cuts. Indeed, we have even
implemented the IAM derived with the inelastic unitarity condition in 
section 3.2.4. We have not found any pole that could hint at the existence of 
such a resonance.

\begin{figure}

\vspace{-0.8cm}

\leftskip -2cm
\begin{center}
\mbox{\epsfysize=6.5cm\epsffile{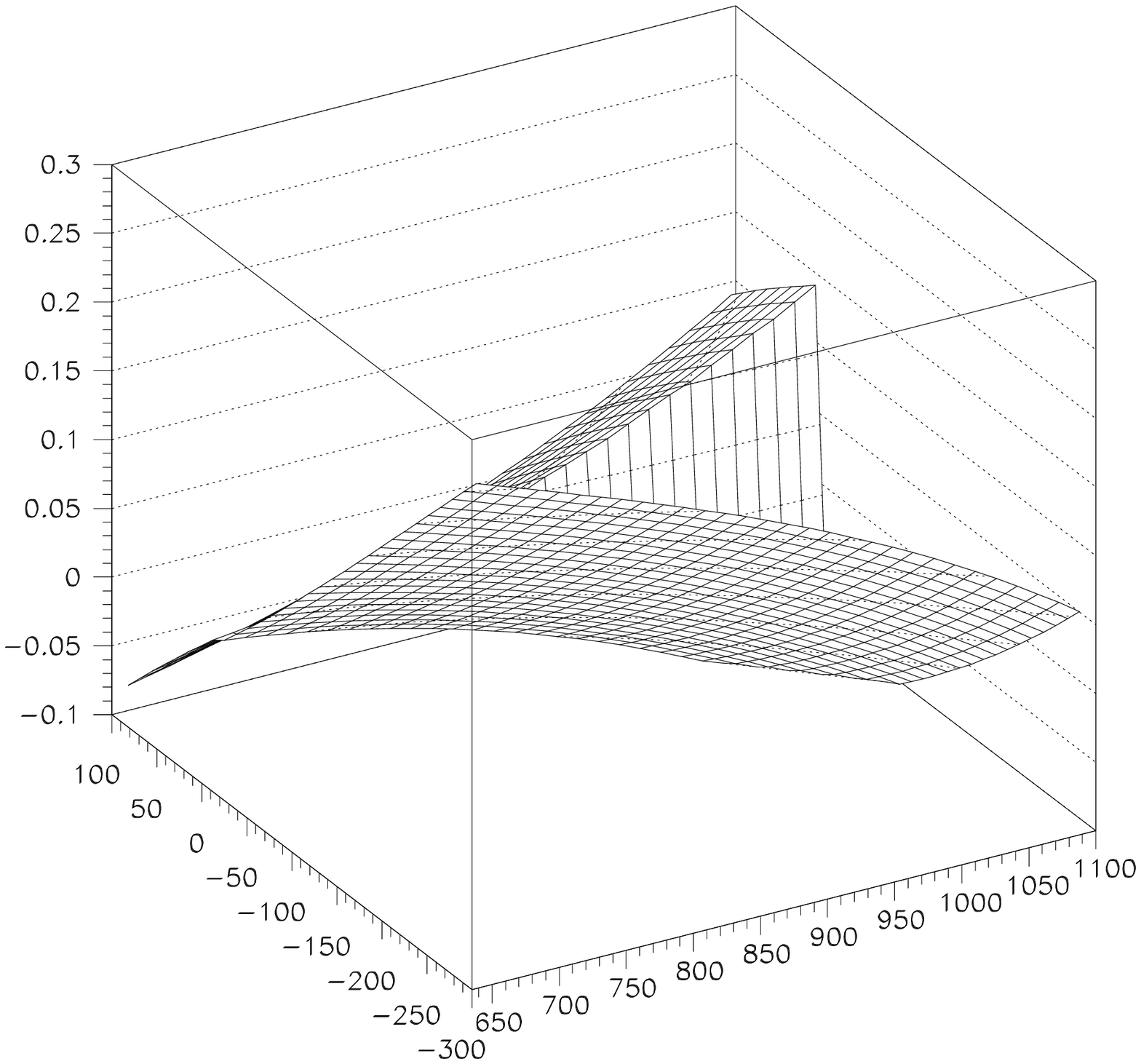}
\epsfysize=6.5cm\epsffile{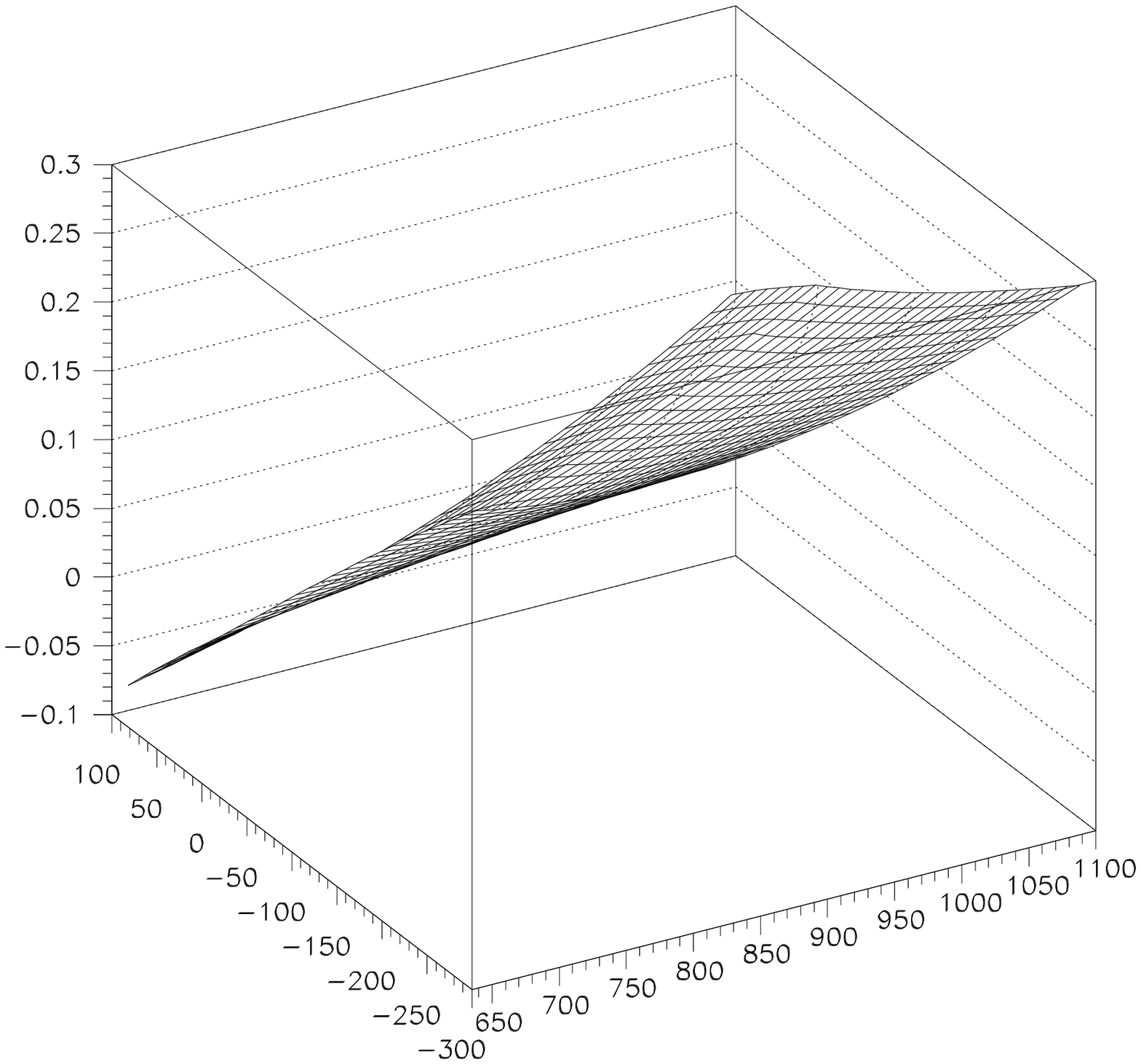}}
\mbox{\epsfysize=6.5cm\epsffile{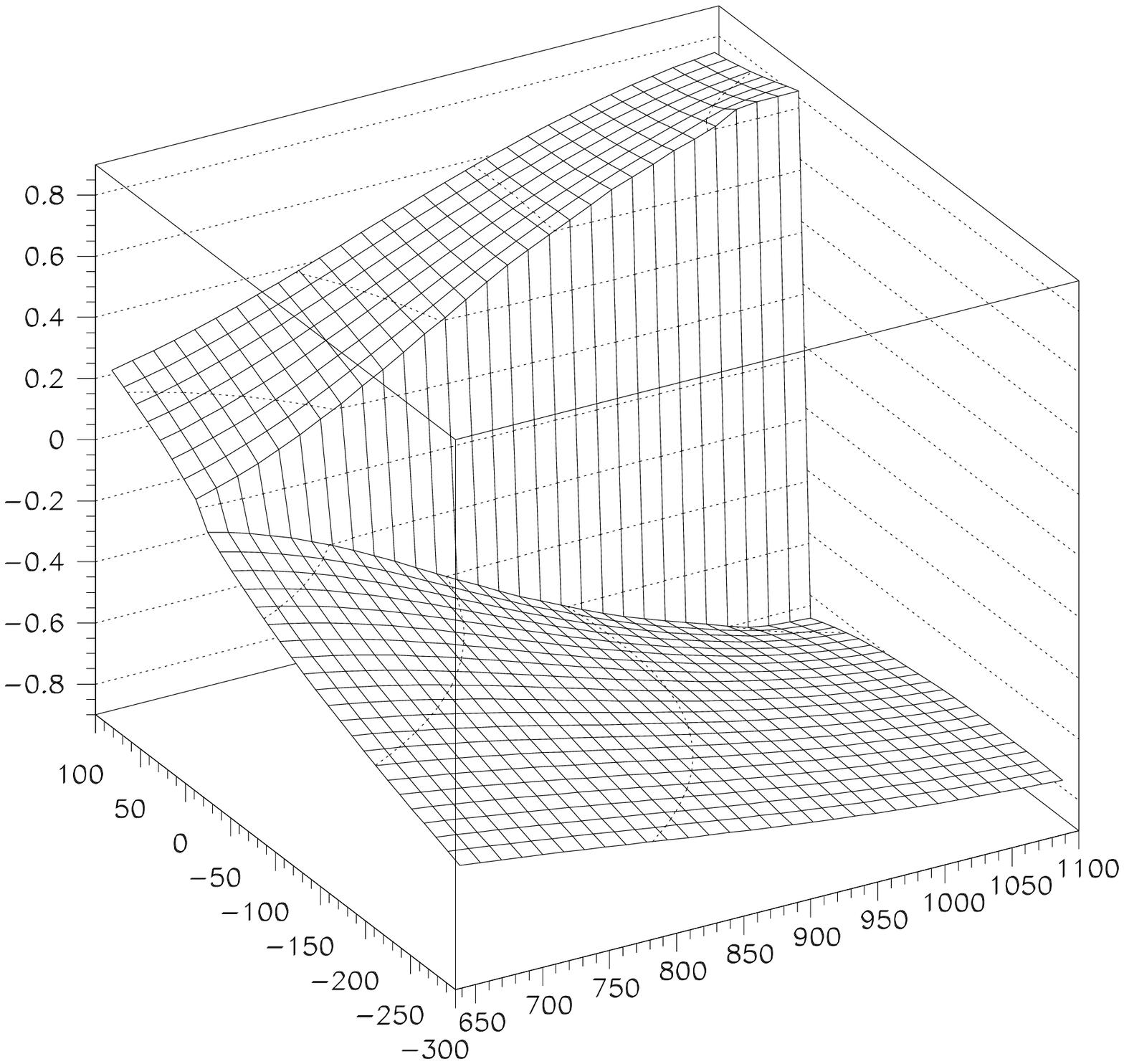}
\epsfysize=6.5cm\epsffile{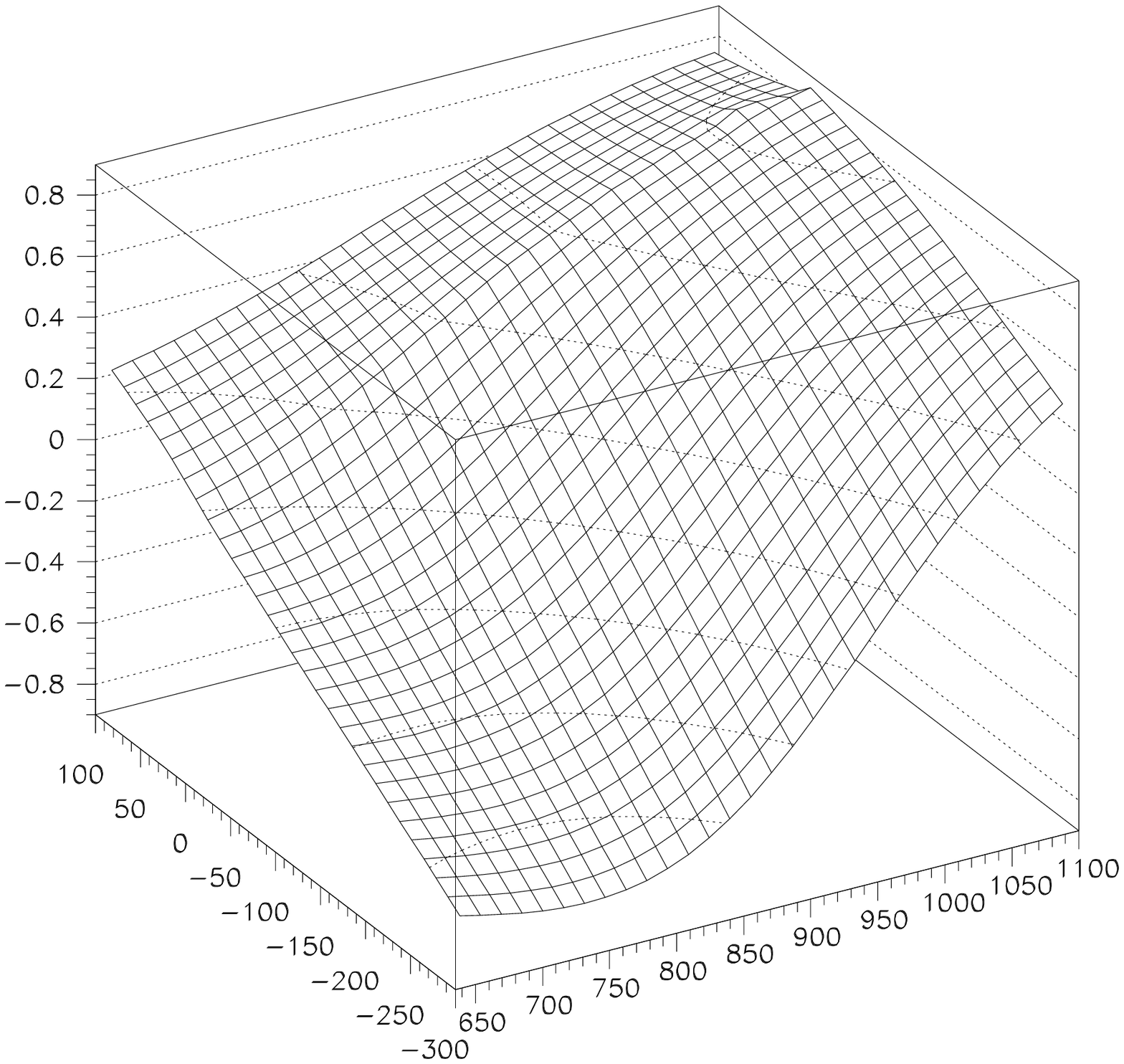}}
\mbox{\epsfysize=6.5cm\epsffile{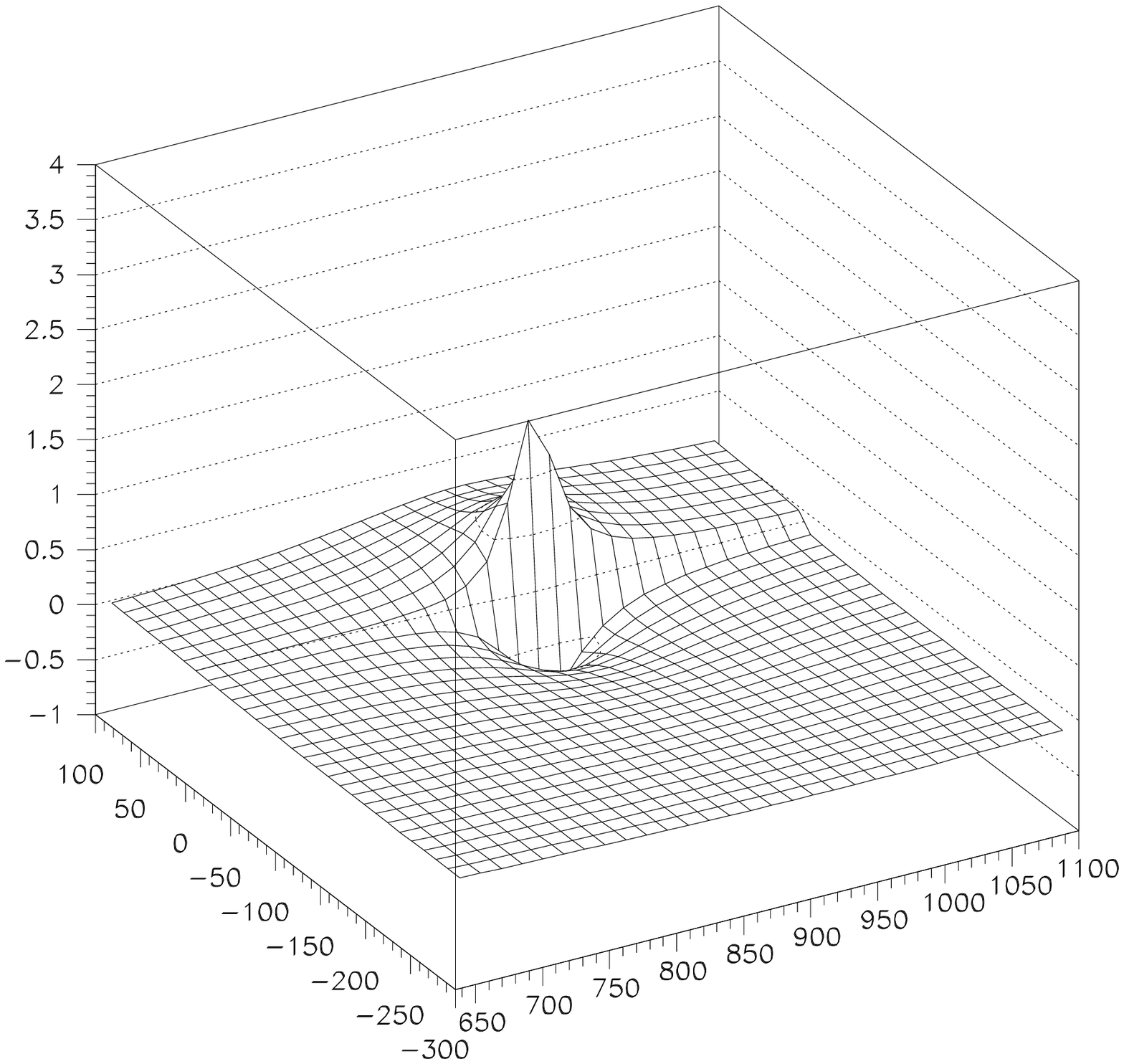}
\epsfysize=6.5cm\epsffile{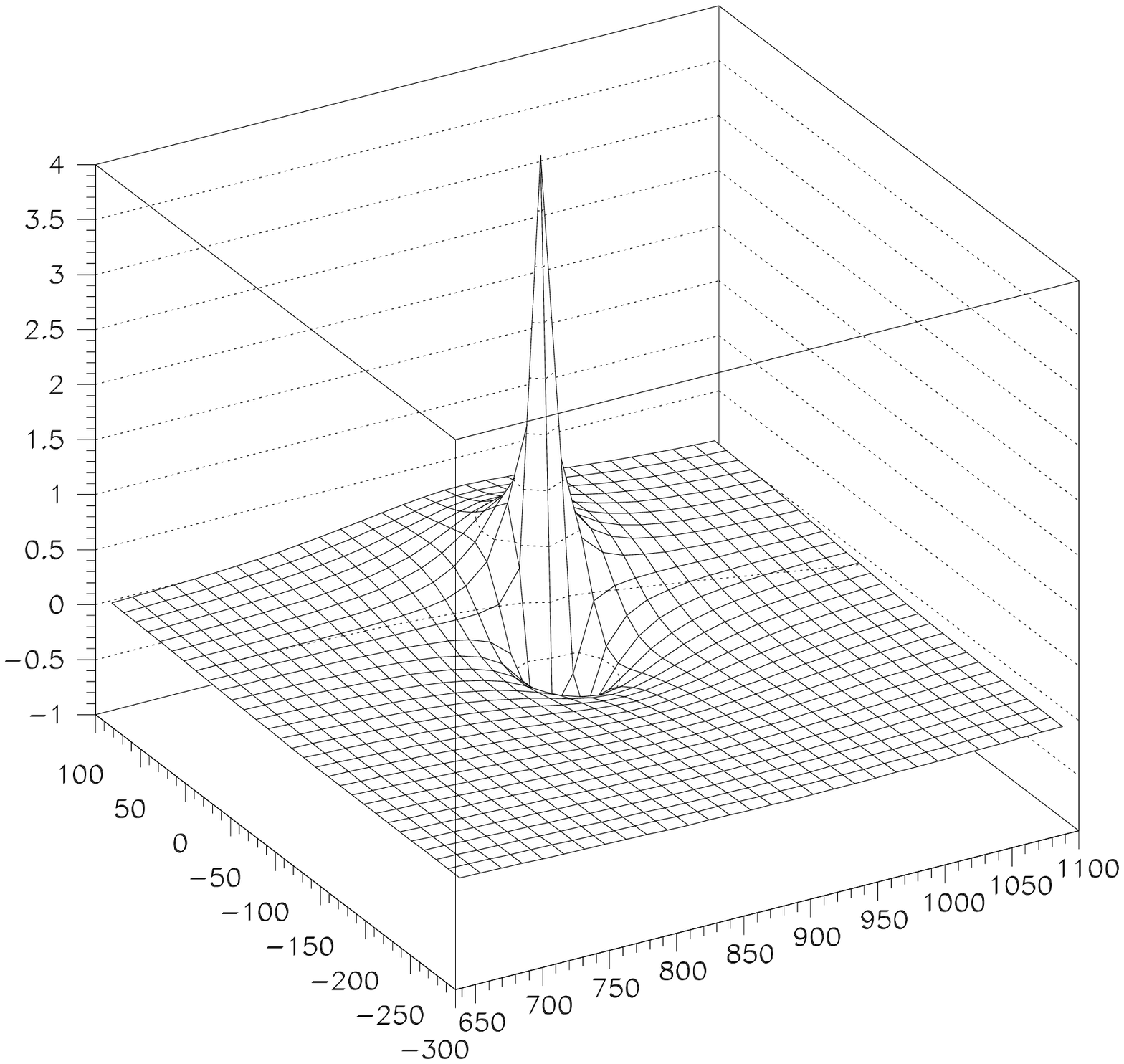}}
\end{center}

\vspace {-.5cm}

\leftskip 1cm
\rightskip 1cm
{\footnotesize {\bf Figure 8.-} Imaginary parts of
the $\pi K \rightarrow \pi K$ IAM amplitudes in the complex $s$ plane.
The first row is the $(I,J)=(3/2,0)$ channel, the second is $(1/2,0)$
and the bottom is $(1/2,1)$. Again, the left plots correspond to the first
Riemann sheet, and those on the right, to the second.}

\leftskip 0.cm
\rightskip 0.cm
\end{figure}

As we have already explained, we should not expect to find anything since
the approach is not able to reproduce properly
the two kaon unitarity cut and consequently neither its associated sheet
 structure. 

Let us now address to the $\pi K$ elastic scattering case. Again, in Fig.8 we
have displayed the imaginary part of the amplitudes for the $(3/2,0),(1/2,0)$
and $(1/2,1)$. Those pictures on the left represent the first Riemann sheet and
those on the right, the second. Once more the existence of a unitarity
cut is clear, but there is also 
the appearance of a pole in the appropriate channel. In particular,
using the third contour plot of Fig.7, we have 
found a pole in $E_R\sim 890-i 20$ MeV, which using Eq.\ref{respar}
yields again the  mass and width for the $K^*(892)$ resonance that we 
gave in Table 5.

\section{Conclusions}

In this work we have shown how the IAM provides a consistent technique
to accommodate resonances.
Indeed, based on its derivation from Dispersion Theory,
we have made a systematic analysis of its applicability,
which is mainly limited by the
existence of two body inelastic thresholds and by the fact that
the tree level approximation vanishes in some channels.

We have found that it is able to
predict the most relevant features of strong elastic scattering,
 once the chiral parameters
are determined from low energy data. Quantitatively, the errors are hard
to estimate, but we have found in all cases
that the mass of the predicted resonances 
fall within approximately
15\% of their actual values. 
We think this fact
gives a sound basis for its application 
in order to obtain at least a qualitative description
of resonances in the strongly interacting
symmetry breaking sector.

Moreover, once
we force the IAM results to fit the actual resonance mass values, we get a
remarkably good fit which is able to reproduce the experimental data
up to the next relevant two body inelastic threshold. Following that 
procedure, we have given the unitarized $SU(2)$ ChPT fit to $\pi\pi\rightarrow\pi\pi$
as well as that of $SU(3)$ to $\pi\pi$ and $\pi K$ elastic scattering.
For the first time we have estimated the values of the unitarized chiral
parameters together with their error bars. 
These values do not lie very far from those obtained without the IAM,
and therefore do not spoil the low energy expansion, as can be noticed from the
scattering lengths that we have given.

With this fit, we have calculated several low-energy phenomenological 
parameters, like the scattering lengths. Our values differ from those obtained at
$\Od(p^4)$ due to the unitarization. However we expect that they include 
other corrections due to unitarity and resonant effects.

We consider that it would not make any sense to try to 
reduce the error bars in the unitarized parameters within this approach.
One has to keep in mind that we have neglected higher order ChPT 
corrections, isospin-breaking contributions, and that we have used high 
energy data which is very sensible to such effects.
It is quite likely that, in order to obtain results consistent
to a higher degree of  accuracy, the IAM in the simple version that has been
used here, will not be enough. 

Finally we have also shown how 
the IAM yields the proper analytic structure in the
complex $s$ plane, in contrast with other unitarization techniques.
Indeed, we have found that the apparent resonant behavior
that is observed on phase shifts, is produced by the corresponding poles
in the second Riemann sheet, meeting the strict requirements imposed by
general relativistic Quantum Field Theory.

Therefore, we think that the IAM and unitarization
by means of dispersion theory is the most natural and 
economic way to extend the applicability of Chiral Lagrangians.
 We have seen
however, that its main limitations come from the existence
of two body inelastic thresholds. Nevertheless, work is still in
progress in the subject, the IAM has been recently
applied to other processes and higher order ChPT calculations
will be soon available. As far as some other physically
relevant features 
do not lie very far from the present applicability limits,
it seems very likely that they can be reproduced in 
the near future.

\section*{Acknowledgments}

J.R.P would like to thank the Theory Group at Berkeley
for their kind hospitality and
the Jaime del Amo Foundation for a fellowship.
He also wishes to thank M.S.Chanowitz for calling
his attention on the $f_0$ problem. We have also profited from 
discussions with U.G. Mei{\ss}ner on the  $\pi K$ 
elastic scattering amplitudes and  M.R.Pennington 
on the Adler zero problem and the left cut contribution.
We thank D.Toublan too, 
for pointing an error in our calculations
as well as C.Carone for reading the manuscript.
This work has been partially supported by the Ministerio de 
Educaci\'on y Ciencia (Spain) (CICYT  AEN93-0776). Partial support
by US DOE under contract DE-AC03-76SF00098 is gratefully acknowledged.

\appendix

\section{Elastic scattering amplitudes in $SU(3)$ ChPT and Unitarity}

The first calculation of elastic $\pi K$ scattering was 
later performed in \cite{Bernard1}. These amplitudes were given in terms of
physical as well as {\it lowest order} masses and decay constants, which are 
usually denoted by $M_P,F_P$ and $M^0_P,F^0$, respectively ($P$ being either
$\pi,K$ or $\eta$).
Of course, the only measurable parameters are the first, and when comparing
with experimental observations, one has to eliminate those from  {\it lowest
order} in terms of the physical constants. 

Indeed, it is possible
to find \cite{GL2} the relation between $M_\pi^0$ and $M_\pi$ as well as that
between $M_K^0$ and $M_K$. Unfortunately, at {\it lowest order} there is only one
$F^0$, which is related both to $F_\pi$ and $F_K$. Hence, whenever one finds
$F^0$ in an expression there are two choices: either relate it to $F_\pi$ 
or to $F_K$. The difference between the two choices will be one order higher
in the chiral expansion.
For instance, if one has an $\Od(p^2)$ expression with $F_0^2$, in principle
one can substitute it by $F_\pi^2,F_K^2$ or $F_\pi\cdot F_K$. All these choices are
equally acceptable. When one is working only with pions, the natural choice
is the one that leaves all the expressions in terms of $F_\pi$. 
When one  is dealing both with pions and kaons, it is not so obvious.
However, once one choice is done for the ${\cal O}(p^2)$ term, we have to 
keep it for the ${\cal O}(p^4)$ contribution, otherwise one would
violate
perturbative unitarity, Eq.\ref{uni}.

Surprisingly, in the amplitude in the
literature \cite{Bernard2}, which is the one we had also followed
in our previous work \cite{UpiK}, the choice for the
${\cal O}(p^2)$ term is different from that of the
${\cal O}(p^4)$ contribution that yields the imaginary part.
Indeed, the ${\cal O}(p^2)$ term is written just in terms of $F_\pi$
whereas $T^U$ is written in terms of $F_\pi$ and $F_K$.
As a consequence, there is a $F_K^2/F_\pi^2$ factor of difference
between $\mbox{Im}t^{(1)}$ and $\sigma_{\pi K} \vert t^{(0)}\vert^2$. 
Numerically that amounts to a
$(1.22)^2\simeq 1.5$ factor.

Thus, we have rederived from the original work \cite{Bernard1} the amplitudes 
in terms of physical quantities, (also correcting some small errata)
 so that they satisfy perturbative 
unitarity. We have chosen to write the formulae symmetrically with respect
to $F_\pi$ and $F_K$. But the other choices are equally acceptable.
The result is:
\begin{eqnarray}
T^{3/2}(s,t,u)&=&\frac{M_\pi^2+M_K^2-s}{2F_\pi F_K}+T^T_4(s,t,u)+T^P_4(s,t,u)+
T^U_4(s,t,u)+\Od(s^3)  \\  \nonumber
T^T_4(s,t,u)&=& \frac{1}{16F_\pi F_K}(M_\pi^2-M_K^2)(3\mu_\pi-2\mu_K
+\mu_{\eta} ) \\   \nonumber
T^P_4(s,t,u)&=& \frac{2}{F^2_\pi F^2_K}\left\{ 4L_1^r(t-2M_\pi^2 )(t-2M_K^2 )
+ 2L_2^r \left[ (s - M_\pi^2-M_K^2)^2 + (u - M_\pi^2-M_K^2)^2 \right]\right.
\\ \nonumber
&+&  L_3^r\left[ (u - M_\pi^2-M_K^2)^2 + (t-2M_\pi^2 )(t-2M_K^2)\right]
+ 4L_4^r\left[t(M_\pi^2+M_K^2) - 4M_\pi^2M_K^2\right]\\ \nonumber
&+&\left. 2L_5^rM_\pi^2(M_\pi^2-M_K^2-s)+8(2L_6^r+L_8^r)M_\pi^2M_K^2 \right\} \\
\nonumber
T^U_4(s,t,u)&=& \frac{1}{4F^2_\pi F^2_K}\left\{ 
\frac{3}{2}\left[ (s-t)\left(L_{\pi K}(u)+L_{K\eta}(u)
-u\left(M_{\pi K}^r(u)+M_{K\eta}^r(u)\right)\right) 
\right.\right. \\ \nonumber
&+& \left.\left(M_K^2 - M_\pi^2)^2(M_{\pi K}^r(u)+M_{K \eta}^r(u)
\right)\right] 
+ t(u-s)[2M_{\pi\pi}^r(t)+M_{KK}^r(t)] \\ \nonumber
&+& \frac{1}{2} (M_K^2 - M_\pi^2) [ K_{\pi K}(u)(5u-2M_K^2-2M_\pi^2)+
K_{K \eta}(u)(3u-2M_K^2-2M_\pi^2)] \\   \nonumber
&+&\frac{1}{8}J_{\pi K}^r(u)\left[ 11u^2 -
12u(M_K^2+M_\pi^2)+4(M_K^2+M_\pi^2)^2] + 
+ J_{\pi K}^r(s)(s-M_K^2-M_\pi^2\right)^2  \\ \nonumber
&+&\frac{3}{8}J_{K \eta}^r(u) \left(u - \frac{3}{2}(M_K^2+M_\pi^2)\right)^2 
+ \frac{1}{2}J_{\pi\pi}^r(t)t(2t-M_\pi^2) +
\frac{3}{4}J_{KK}^r(t)t^2 \\ \nonumber
&+&\left. \frac{1}{2}J_{\eta\eta}^r(t)M_\pi^2
\left(t-\frac{8}{9}M_K^2\right)\right\}\\ \nonumber
\end{eqnarray}
The functions $M_{PQ},L_{PQ},K_{PQ},J_{PQ},\mu_P$, with $P,Q=\pi,K,\eta$,
 can be found in \cite{GL2}
although they should be written in terms of physical quantities.

We have verified analytically that this amplitude satisfies the 
perturbative unitarity constraint. Moreover, we have used that constraint
as a check of our programs. 

We want to remark again that 
this way to write the $\pi K$ amplitude 
is one of several possible choices, since we could have
chosen to write everything just in terms of $F_\pi$, for example.
The important point is to keep the same choice
{\em both} for the ${\cal O}(p^2)$ and
the ${\cal O}(p^4)$.

For completeness, we will also give the $SU(3)$ formulae used in this
work for $\pi\pi$ scattering, because they have also appeared with 
some minor errata in the literature:
\begin{eqnarray}
A(s,t,u)&=&\frac{(s-M_{\pi}^2)}{F_{\pi}^2}+B(s,t,u)+C(s,t,u)+\Od(s^3)\\ \nonumber
B(s,t,u)&=&\frac{1}{F_{\pi}^4} \left\{
\frac {M_{\pi}^4}{18} J_{\eta\eta}^r (s) +
\frac{1}{2} ( s^2 - M^4_{\pi}) J_{\pi\pi}^r(s) + 
\frac {1}{8} s^2 J_{KK}^r (s) \right. \\ \nonumber
 &+& \left. \frac{1}{4}( t - 2 M_{\pi}^2)^2 J_{\pi\pi}^r (t) + t(s-u) \left[
M_{\pi\pi}^r (t)+\frac{1}{2} M_{KK}^r (t) \right]+ (t \leftrightarrow u) 
\right\}\\
\nonumber
  C(s,t,u)&=&\frac{4}{F_{\pi}^4} \left\{ (2L_1^r + L_3 )( s - 2M_{\pi}^2)^2 +
 L_2^r [ (t-2M_{\pi}^2)^2 + ( u - 2M_{\pi}^2)^2] +\right. \\ \nonumber
 &+& \left. (4L_4^r+2L_5^r)M_{\pi}^2(s-2M_{\pi}^2) + 
(8L_6^r+4L_8^r)M_{\pi}^4 \right\} \\ \nonumber
\end{eqnarray}


\end{document}